\documentclass[a4paper,fleqn,usenatbib]{mnras}

\usepackage{newtxtext,newtxmath}

\usepackage[T1]{fontenc}
\usepackage{ae,aecompl}

\usepackage{graphicx}	
\usepackage{amsmath}	
\usepackage{amssymb}	

\usepackage{lscape}
\usepackage{subfigure}
\usepackage{multirow}
\usepackage{url}
\usepackage{threeparttable}
\usepackage{natbib, aas_macros}
\title[Fe-K lags produced by clouds]{
X-ray short-time lags in the Fe-K energy band produced by scattering clouds in active galactic nuclei
}

\author[M.\,Mizumoto et al.]{
Misaki Mizumoto$^{1,2}$\thanks{E-mail: mizumoto@astro.isas.jaxa.jp, mizumoto.misaki@gmail.com (MM)},
Chris Done$^{3}$,
Kouichi Hagino$^{4}$,
Ken Ebisawa$^{1,2}$,
\newauthor
Masahiro Tsujimoto$^{1}$ \&
Hirokazu Odaka$^{5}$
\\
$^1$Institute of Space and Astronautical Science (ISAS), Japan Aerospace Exploration Agency (JAXA), 3-1-1 Yoshinodai, Chuo-ku,\\
Sagamihara, Kanagawa, 252-5210, Japan\\
$^2$Department of Astronomy, Graduate School of Science, The University of Tokyo, 7-3-1 Hongo, Bunkyo-ku, Tokyo, 113-0033, Japan\\
$^3$Centre for Extragalactic Astronomy, Department of Physics, University of Durham, South Road, Durham DH1 3LE, UK\\
$^4$Department of Physics, Faculty of Science and Technology, Tokyo University of Science, 2641 Yamazaki, Noda, Chiba, 278-8510, Japan\\
$^5$Kavli Institute for Particle Astrophysics and Cosmology, Stanford University, 452 Lomita Mall, Stanford, CA 94305, USA
}

\date{Accepted XXX. Received YYY; in original form ZZZ}

\pubyear{2018}

\begin{document}
\label{firstpage}
\pagerange{\pageref{firstpage}--\pageref{lastpage}}
\maketitle

\begin{abstract}

  X-rays illuminating the accretion disc in active galactic nuclei
  give rise to an iron K line and its associated reflection
  spectrum which are lagged behind the continuum variability by the
  light-travel time from the source to the disc.  The measured lag
  timescales in the iron band can be as short as $\sim R_g/c$,
  where $R_g$ is the gravitational radius, which is often interpreted
  as evidence for a very small continuum source close to the event horizon
  of a rapidly spinning black hole.  However, the short lags can
  also be produced by reflection from more distant material,
  because the primary photons with no time-delay 
  dilute the time-lags caused by the reprocessed photons.  
  We perform a Monte-Carlo simulation to
  calculate the dilution effect in the X-ray reverberation lags
  from a half-shell of neutral material placed at $100\,R_g$ from the
  central source.  This gives lags of $\sim2\,R_g/c$, but the
  iron line is a distinctly narrow feature in the lag-energy plot, whereas
  the data often show a broader line.  We show that both the short lag
  and the line broadening can be reproduced if the scattering material
  is outflowing at $\sim0.1c$. The velocity structure in the wind can
  also give shifts in the line profile in the lag-energy plot
  calculated at different frequencies.  Hence we propose that the
  observed broad iron reverberation lags and shifts in profile
  as a function of frequency of variability can arise from a disc wind
  at fairly large distances from the X-ray source.

\end{abstract}




\section{Introduction} \label{sec1}

Reverberation lags in AGNs give important clues about nature and
geometry of the material around the central super-massive black hole.
The intrinsic hard X-ray corona illuminates the material, producing an
iron K (Fe-K) emission line and continuum reflection which includes
strong soft X-ray emission if the material is partially ionised. 
The reprocessed emission lags behind the primary emission
due to the light-travel distance.
This kind of lag was first significantly detected in {\it
  XMM-Newton} observations of 1H 0707--495 \citep{fab09}, in which the
light curve of the fast variability in the soft X-ray band
(0.3--1~keV) was lagged by $\sim 30$~s behind the same variability
seen in a hard band (1--4~keV).  This corresponds to $\sim3\,R_g/c$
for the black-hole mass of 1H 0707--495
($M_\mathrm{BH}=2\times10^6M_\odot$; \citealt{zho05}), where
$R_g=GM_\mathrm{BH}/c^2$ and $M_\mathrm{BH}$ is the black hole mass.
Lags were also detected in this object in the energy band
containing the Fe-K line \citep{kar13b}, such that photons in the
Fe-K energy band lag behind those in the adjacent energy bands by
$\simeq50$~s, corresponding to $\sim5\,R_g/c$.  The full lag-energy
plot for the fast variability in 1H 0707--495 shows that the lags
associated with the Fe-K line emission are significant across the
entire 5--7~keV band, so this is a broad feature.  To date, there are
$\sim20$ AGNs where such broad Fe-K reverberation lags can be measured, with
amplitudes of $1-9\,R_g/c$, at frequencies of order  
$10^{-4} (10^7M_\odot/M_\mathrm{BH})$~Hz
\citep{kar16}.

Both the observed short lag times and the broad Fe-K feature can be
explained by extreme relativistic disc-reflection around a high spin
Kerr black hole (\citealt{fab09,kar13b}).  In this scenario,
a very compact X-ray corona just above the black hole (lamppost
geometry) illuminates the innermost regions of the disc.  The
light-travel time from the corona to the disc is several $R_g/c$,
explaining the observed small lag times, while the Fe-K line in the
reflected spectrum is broadened and skewed due to the fast velocities
and strong gravity experienced by the disc material close to the black
hole (e.g., \citealt{zog11,kar13b,kar13,kar14,cac14,wil16,cha16}).

However, the observed lag amplitude underestimates the light-travel
distance due to the ``dilution'' effect.  The iron line band contains
continuum photons as well as the lagged line emission, and these
correlate with the continuum variability in a reference band with no
lag.  The observed lag is then given by the intrinsic lag multiplied
by the fraction of photons in the band lagged by this light-travel
time.  Therefore, the observed short lag timescales do not necessarily
indicate that the reprocessing matter is close.  The short lags
in 1H 0707--495 and other AGNs can also be explained by distant
clouds at $\gtrsim1000$~light-seconds, corresponding to
$\gtrsim100-600\,R_g/c$ \citep{mil10a,mil10b}.  
\citet{tur17} also
proposed that distant materials at $\sim100\,R_g$ explain the
lag-frequency plot of NGC 4051.  The short lags can then be
interpreted in a very different geometry, one where there is
substantial material above the disc at $\sim100\,R_g$ from the black
hole, plausibly arising from a wind.  This is especially attractive
for 1H 0707--495 as its multi-wavelength spectrum clearly implies that
the accretion flow is super-Eddington \citep{don16}.  This object
should power strong winds, and emission/absorption from this wind can
fit the strong and broad Fe-K emission features observed in the
spectrum without requiring extreme relativistic effects and extreme
super-solar iron abundances \citep{hag16}.

While \citet{mil10a,mil10b} and \citet{tur17} showed 
that lags from a cloud at $\sim100\,R_g$ could explain the
lag-frequency plots, they did not explain the observed Fe-K broad
features in the lag-energy plots.  Thus, in this paper, we investigate
the energy dependence of the lag amplitudes created by more distant
scattering materials.  We examine if both the observed short lag
timescales (corresponding to $\sim$several $R_g/c$) and the broad Fe-K
features can be explained in this model.  We use a Monte-Carlo
simulation, as this is most suitable to quantitatively study the
energy-dependent lag features produced by reverberation in arbitrary
geometries.  First, we explain the method of
Monte-Carlo simulations and our assumed geometry in \S\ref{sec2}, and
show the resultant spectra and lag features in \S\ref{sec3}.  Then we
discuss whether distant reflection can explain the observed short lag
amplitude and the broad Fe-K lag feature in \S\ref{sec4}, and finally
give our conclusions in \S\ref{sec5}.

In this paper, we assume that (1) the scattering material is neutral
rather than ionised, (2) its velocity structure is only radial rather
than including azimuthal components, and (3) the material is
homogeneous rather than clumpy.  Also, we take only reverberation lags
into consideration, rather than including propagation lags which could
be intrinsic to a stratified accretion flow close to the black hole 
(see e.g., \citealt{kot01,are06}).  All these processes are expected
to contribute to shaping the observed lags (see e.g.,
\citealt{gar14,gar15}), and we will extend our work to include them in
subsequent papers.

\section{Method} \label{sec2}

We use a Monte-Carlo simulation code, MONACO 
(MONte Carlo simulation for Astrophysics and COsmology; \citealt{oda11}), 
which is a general-purpose framework for synthesizing X-ray radiation 
from astrophysical objects. 
This code employs the Geant4 toolkit library \citep{ago03,all06} 
for photon transport in complicated geometries, but with 
physical processes for the interaction of photons with matter based 
on the original implementation provided by \citet{oda11}. 
In this work, we consider photoelectric effects followed by 
fluorescence emission, together with all scattering by electrons
(i.e., Compton, Raman, and Rayleigh scatterings). 
These calculations can handle multiple scatterings and 
Doppler effects from velocity structure in the material.
General relativistic effects are not taken into account because the scattering medium is assumed to be located far from the central BH.

\begin{figure}
\includegraphics[width=\columnwidth]{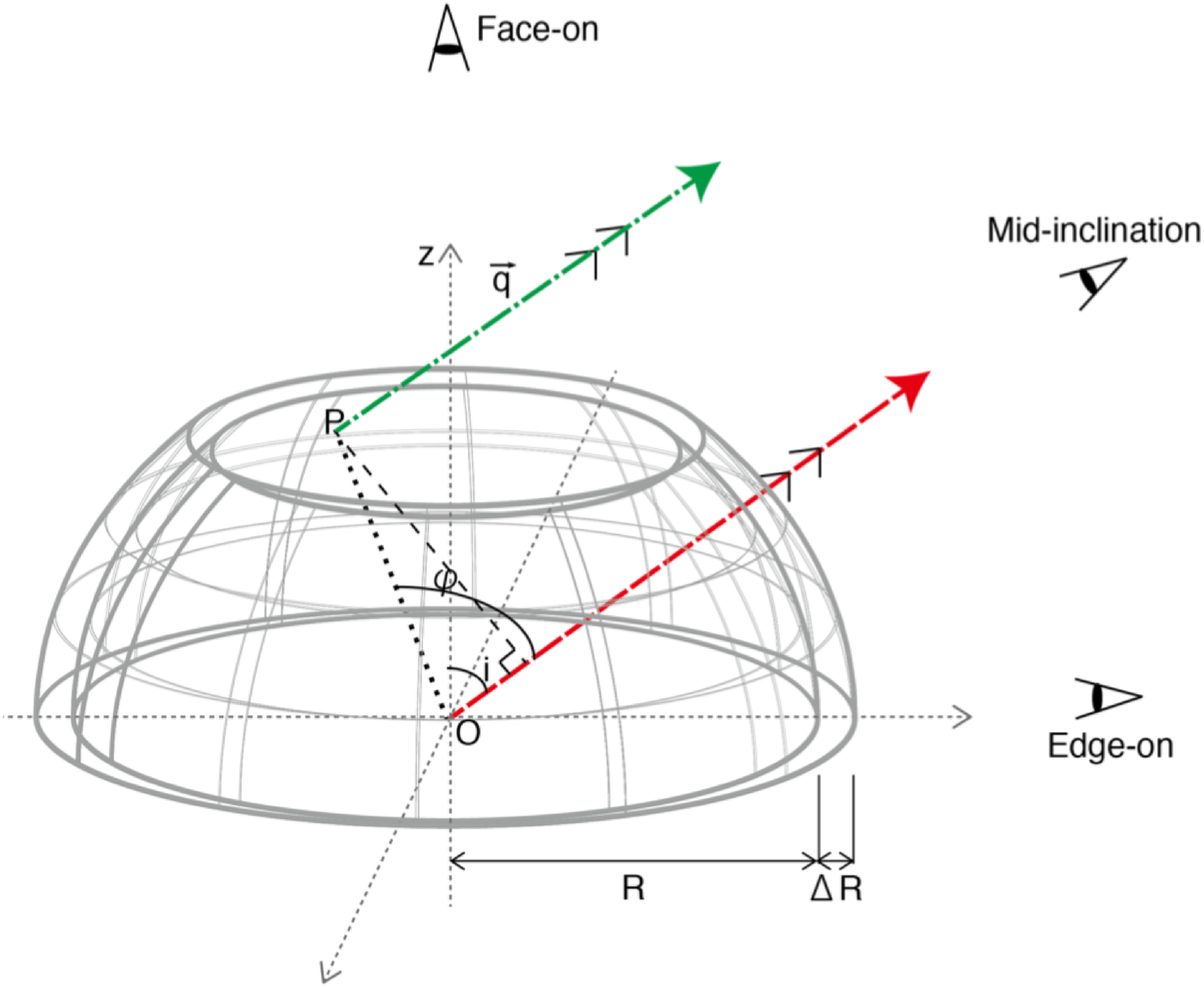}
\caption{
Assumed geometry of the ambient scattering medium and photon paths in our simulations.
The scattering medium (grey solid line) is a ring-shaped part of the upper-half of
a spherical shell, whose thickness ($\Delta R$) is $0.1R$.
The point $P$ is the location where the photon interacts last in the shell, and
$\vec{q}$ is a velocity vector the photon has.
$i$ is the inclination angle and $\phi$ is the scattering angle.
The red dashed/green dot-dashed lines show paths of the
primary/reprocessed photons toward the line-of-sight.
}
\label{fig:geometry}
\end{figure}

We assume a simplified geometry, in which photons are emitted radially from a central point source 
(though in practice, this should have finite extent, which we assume is $5-10\,R_g$).  
We assume that the scattering medium is a section of a spherical shell (Fig.\ \ref{fig:geometry}).
The shell radius is from $R=100\,R_g$ to $110\,R_g$ across zenith angles $45^\circ$--$90^\circ$ 
(i.e., $\Omega/2\pi=0.7$).  
Photons emitted downwards are assumed to be blocked by the accretion disc. 
These will give an additional contribution to the lagged emission but here 
we are interested in the signatures of larger scale reflecting material so we neglect this. 
We assume $M_\mathrm{BH}=10^7\,M_\odot$, 
which means that the light-travel time to the shell is 5000~s.
We note that the simulational results simply scale with $M_{\rm BH}$.

We consider two cases for the shell dynamics. 
We first explore what happens with a static shell,
and secondly assume that the shell is formed by a wind. 
Typically, winds have outflow velocity similar to the escape velocity from its launch point,
i.e.\ $0.14c = 42,000$~km~s$^{-1}$ for a launch radius of $100R_g$
\citep{tom11}. We assume that the turbulent velocity 
is $1000$~km~s$^{-1}$ \citep{hag15}.  
We set the hydrogen radial column density ($N_\mathrm{H}$) of the shell as
$2\times10^{23}\,\mathrm{cm}^{-2}$.  
Incident photons have a power-law spectrum whose photon index is 2. The
number of input photons is $7\times10^8$ in each case, injected as a 
delta function in time from 2--50~keV.
Time-lags in the soft energy band ($<2$~keV) is also important (e.g., \citealt{fab09}), but
we do not treat this band because we now focus on the Fe-K feature around 5--8~keV and most of the soft photons are absorbed by the shell for our assumption of neutral material.

Fig.\,\ref{fig:geometry} shows the assumed geometry. The 
red dashed line shows the path of primary 
X-ray photons which pass through the shell on the line of sight to the
observer. These have unit vector is $\vec{q}$.  Other photons 
emitted along this path are absorbed or scattered in the shell and do not reach us.  
The green dot-dashed line shows the path of a reprocessed photon which
does reach the observer. 
The point $P$ is the location where the photon interacts last in the shell at a time $t_\mathrm{P}$ 
after being emitted from the central source, 
no matter how many times it has been scattered previously. 
This photon has a time-delay of
$t_\mathrm{P} - (\vec{P}\cdot\vec{q})/c$.  
Since MONACO keeps track of $t_\mathrm{P}$, $\vec{P}$ and $\vec{q}$, 
we can calculate the time-delay of each scattered photon relative to the direct photons 
as well as a scattering angle $\phi \equiv \arccos(\vec{p}\cdot\vec{q})$, where
$\vec{p}$ is a unit vector of $\vec{P}$.

\begin{figure*}
 \begin{center}
\subfigure{
\resizebox{8cm}{!}{\includegraphics[angle=270]{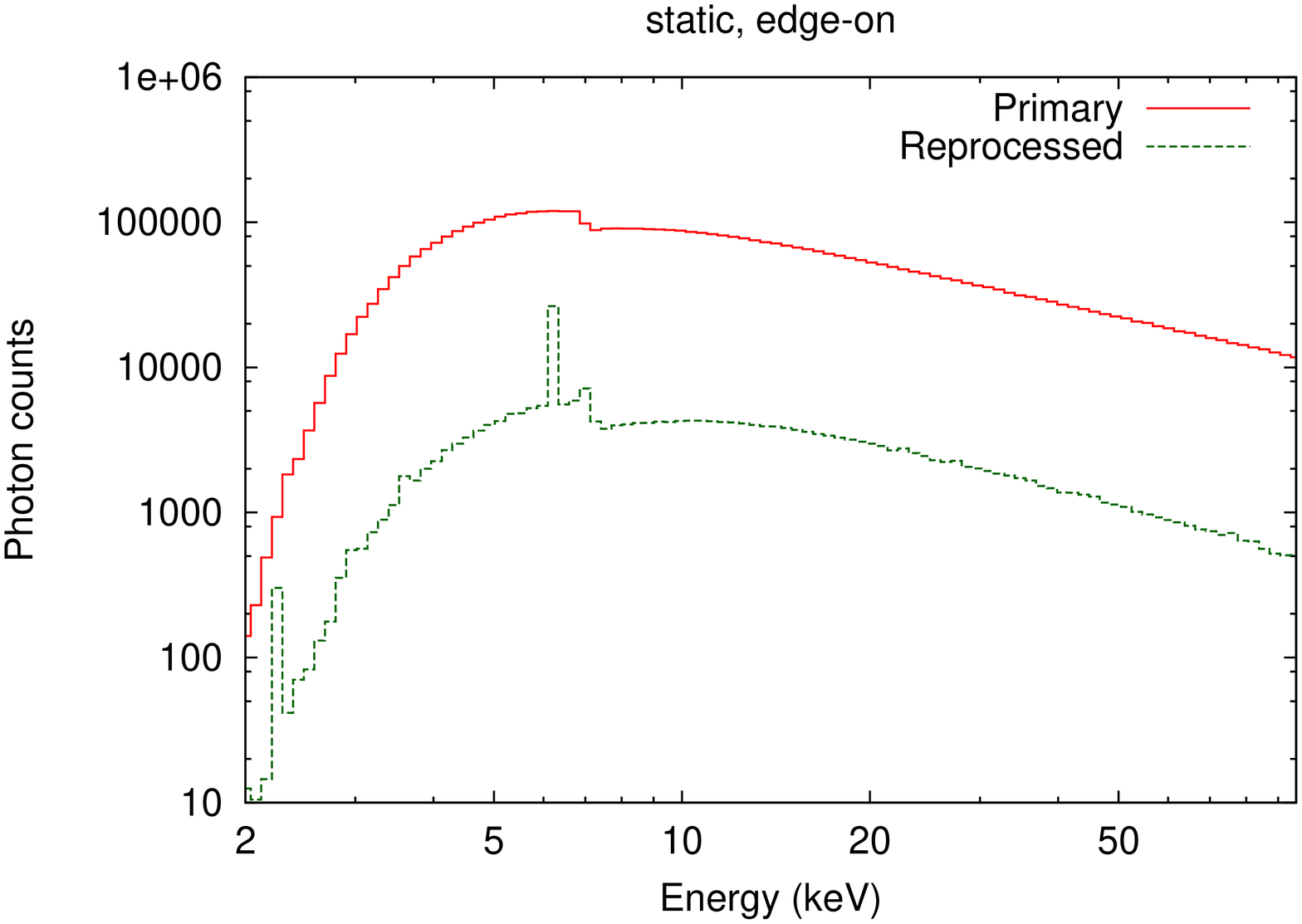}}
\resizebox{8cm}{!}{\includegraphics[angle=270]{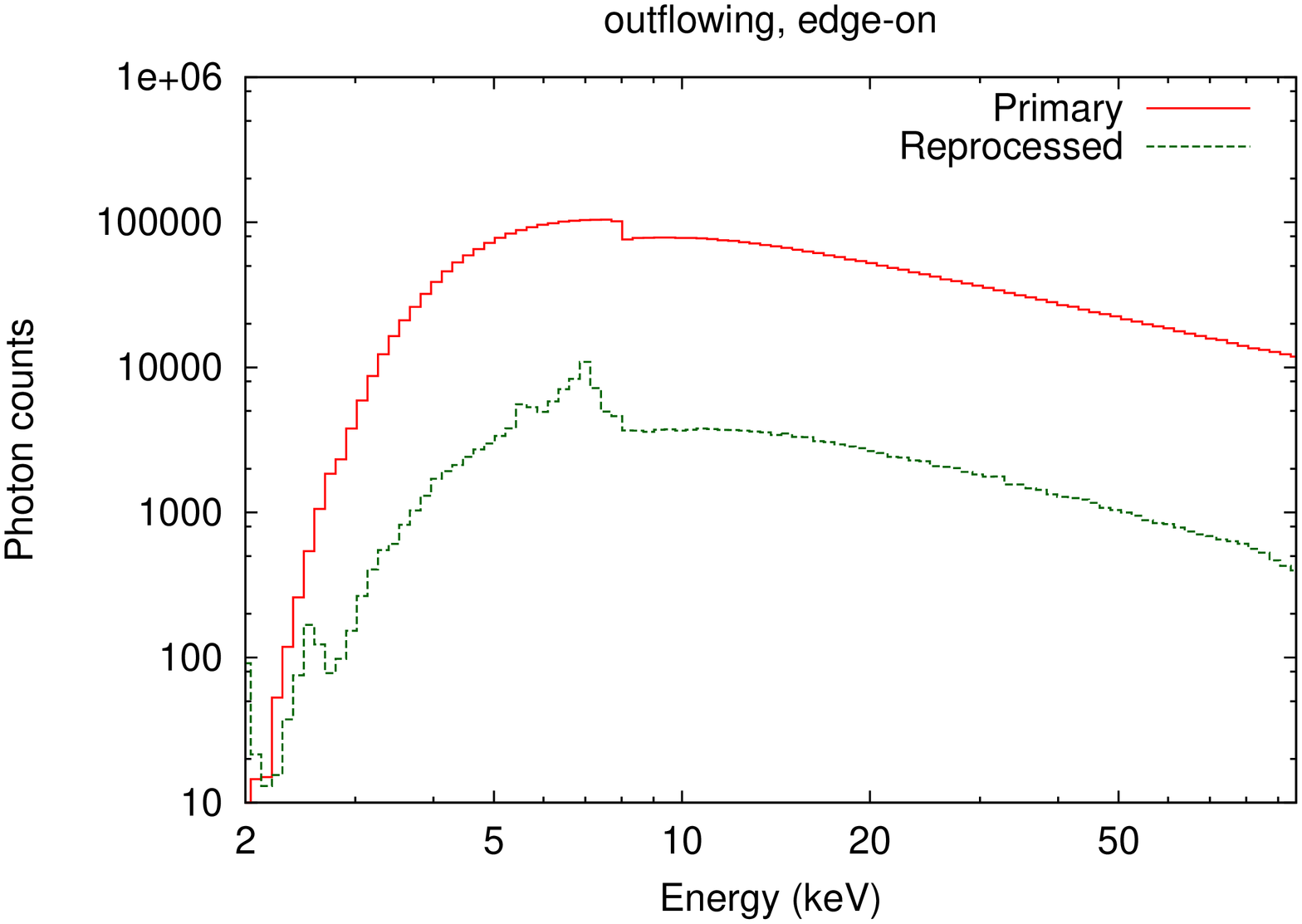}}
}
\subfigure{
\resizebox{8cm}{!}{\includegraphics[angle=270]{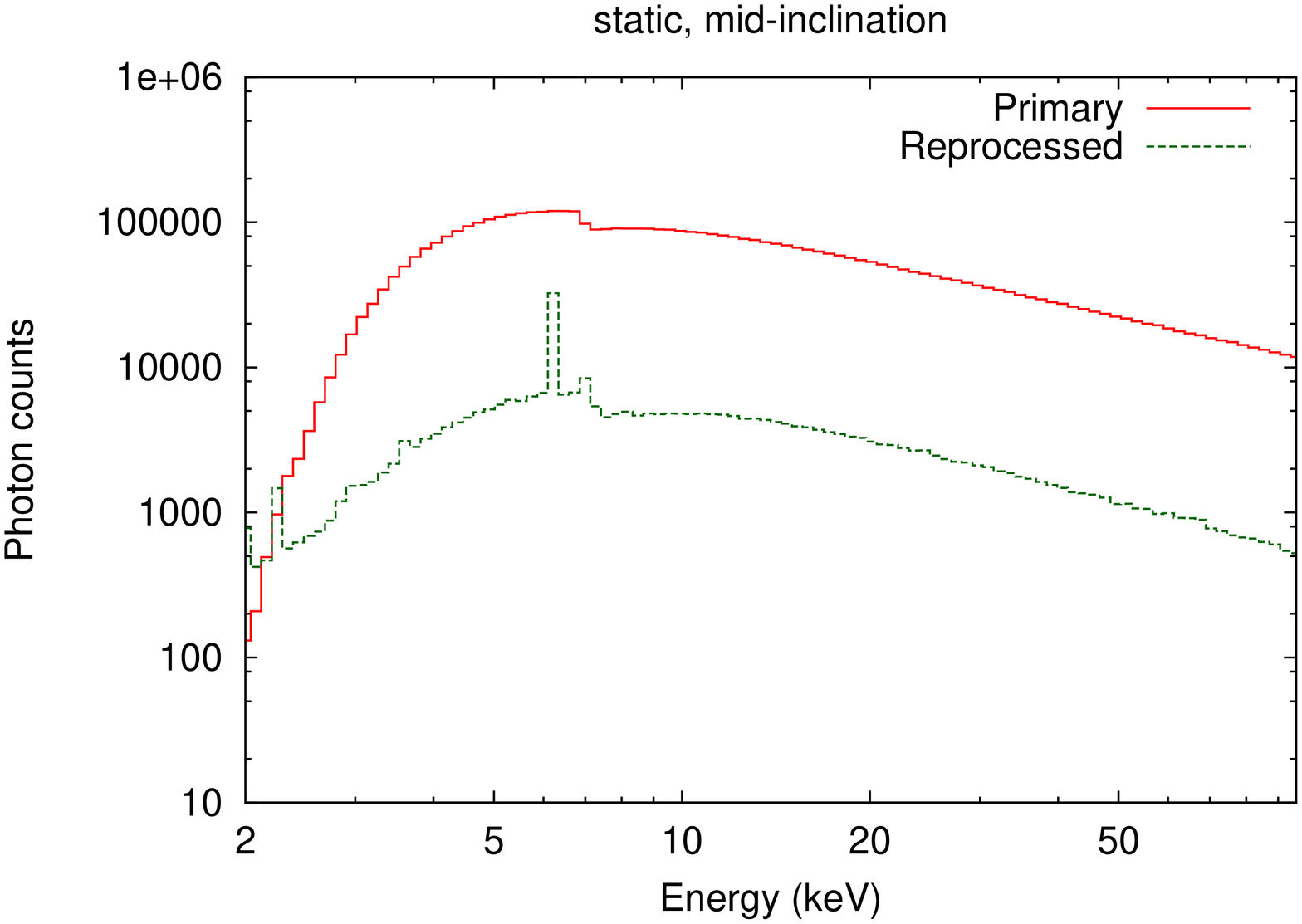}}
\resizebox{8cm}{!}{\includegraphics[angle=270]{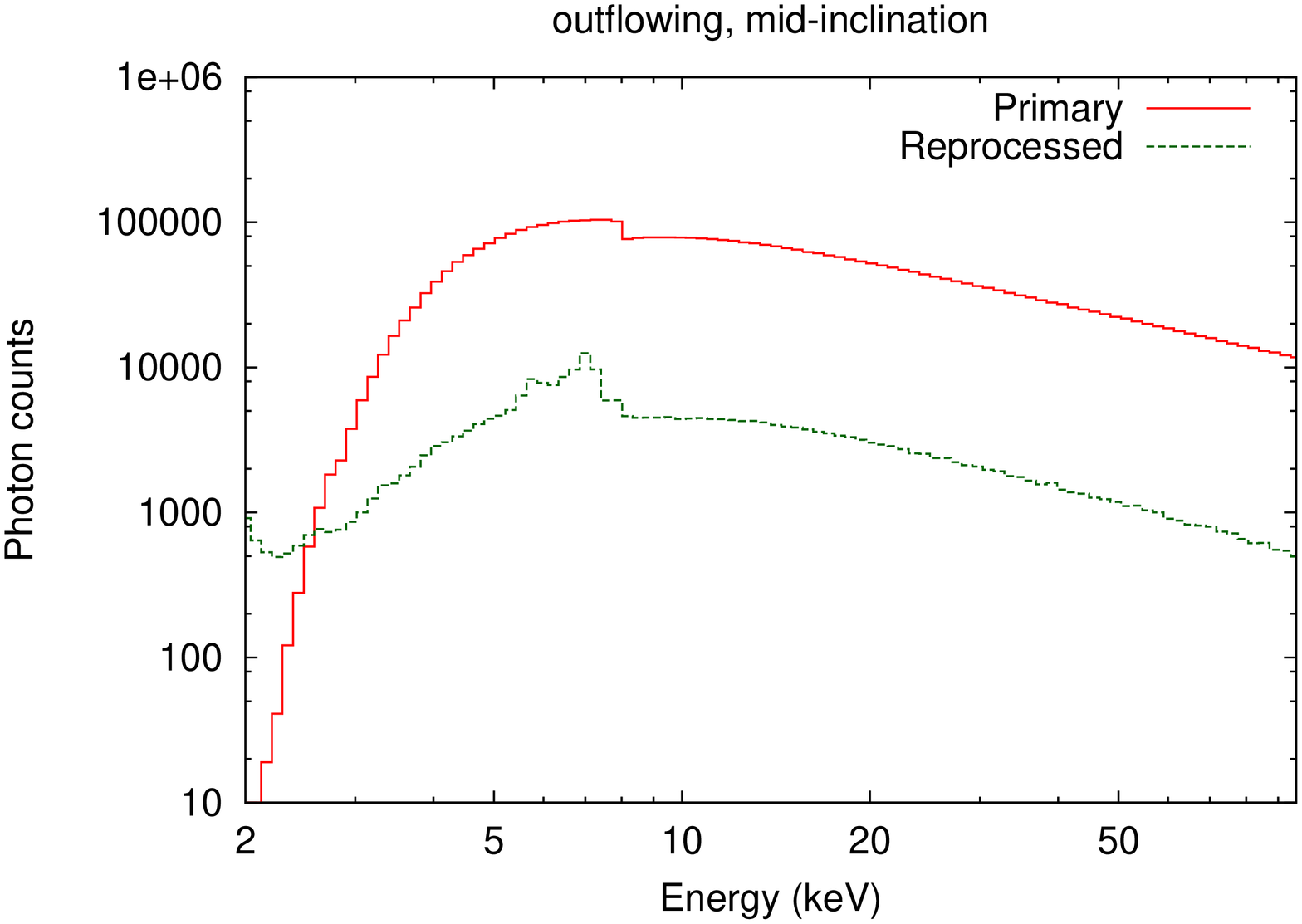}}
}
\subfigure{
\resizebox{8cm}{!}{\includegraphics[angle=270]{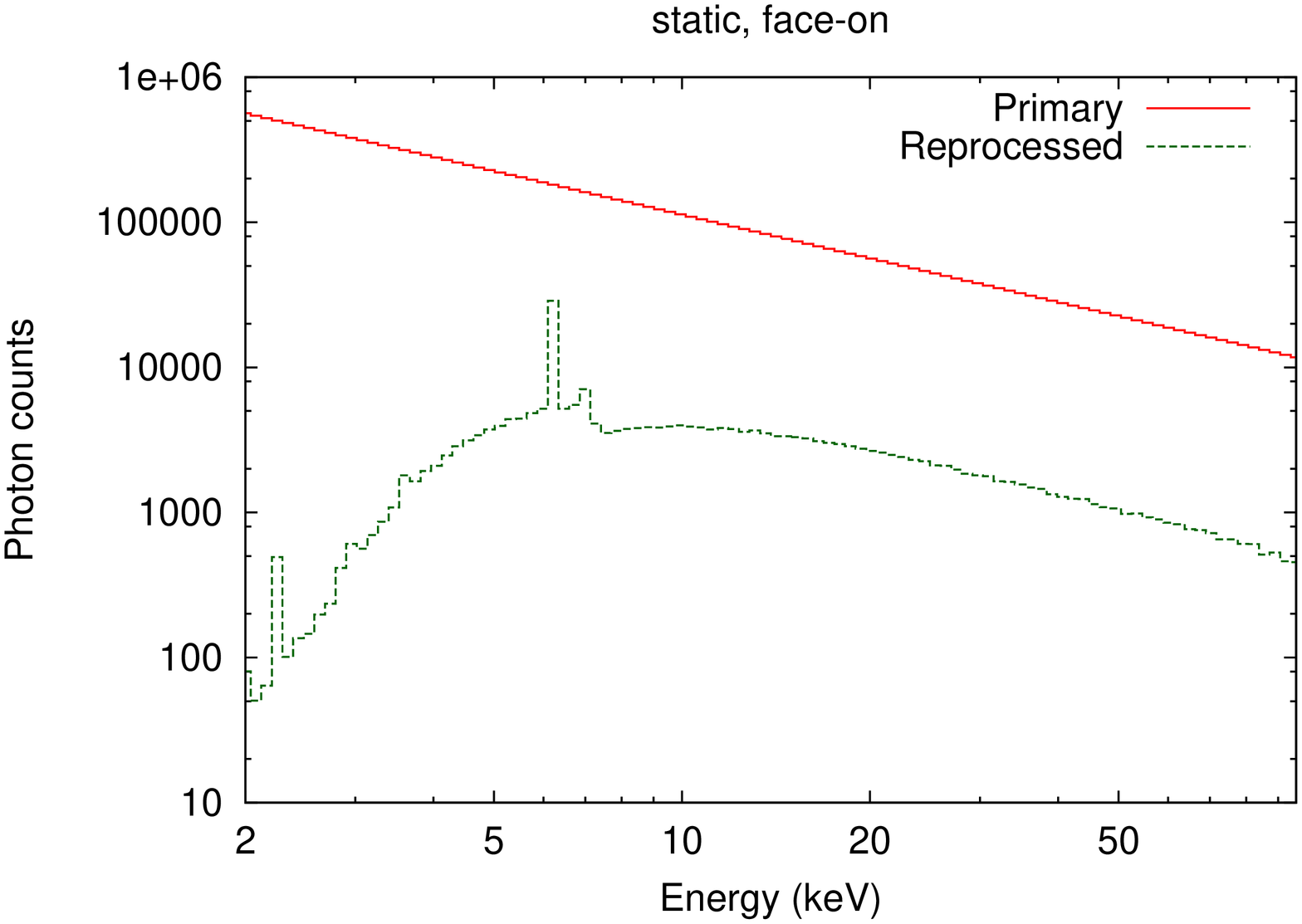}}
\resizebox{8cm}{!}{\includegraphics[angle=270]{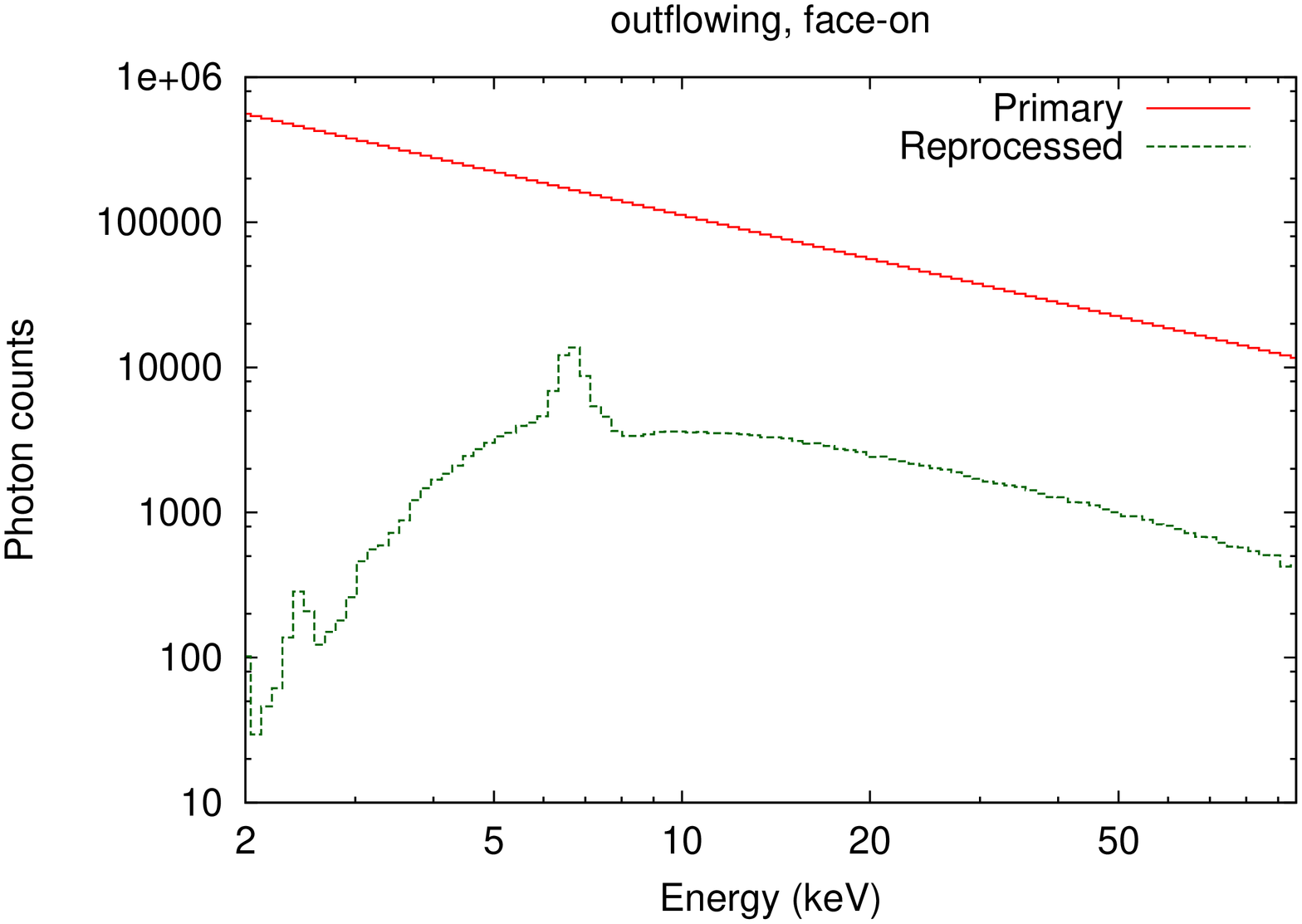}}
}
 \caption{Energy spectra of primary/reprocessed components with different inclinations
for each of the static shell (left) and the outflowing shell (right). 
Each energy-bin-width is made logarithmically equal ($\Delta\log E\,({\rm keV})=0.017$).
Note that the reprocessed photons are dominant below $\sim3$~keV
in the outflowing mid-inclination case,
which significantly affects the energy/frequency dependent lag features.
 }\label{fig:spectrum}
 \end{center}
\end{figure*}

In this paper we assume a fixed size scale and the
  geometry as shown in Fig.\ \ref{fig:geometry}. However, it is fairly simple to assess
  the impact of these parameter choices.  The reprocessed fraction is set by
  the total solid angle of the reprocessing material and its optical
  depth, $\tau = n\sigma_{\rm T} \Delta R$, so that
\begin{equation}
f_{\rm rep}\sim\frac{\Omega}{2\pi}\tau 
\propto (1-\cos\theta_{\rm op} )(n\Delta R),
\end{equation}
where $\theta_{\rm op}$ is the half opening angle of the torus and the
total column density in the wind is $N_{\rm H}=n\Delta R$.  Thus changing
the geometry from a spherical shell to a cylinder makes very little
difference as long as $\theta_{\rm op}$ stays constant. The
scattered fraction scales with 
total column density in the wind, so the fraction of lagged photons
will depend linearly on $N_{\rm H}$. 
The delay-time ranges
are from $\min\{0,R/c\,[1-\sin(\theta_{\rm op}+i)] \}$ (near side of
the wind to the observer) to
$R/c\,(1+\sin i)$ (far side) where $i$ is the zenith angle, so these depend
linearly on the scale of the scatter. Thus the results shown here can
be easily scaled to give an approximate result for changes in $R$,
$\theta_{\rm op}$, and $n\Delta R$.

We show results along three
different inclination angles, collecting photons in  $\cos i$ ranges
between 14/15 and 15/15
(face-on), between 7/15 and 8/15 (mid-inclination, which just
intercepts the wind for our assumed $\theta_{\rm op}=45^\circ$), and between 0/15
and 1/15 (edge-on).

\begin{figure*}
 \begin{center}
\subfigure{
\resizebox{8cm}{!}{\includegraphics[angle=270]{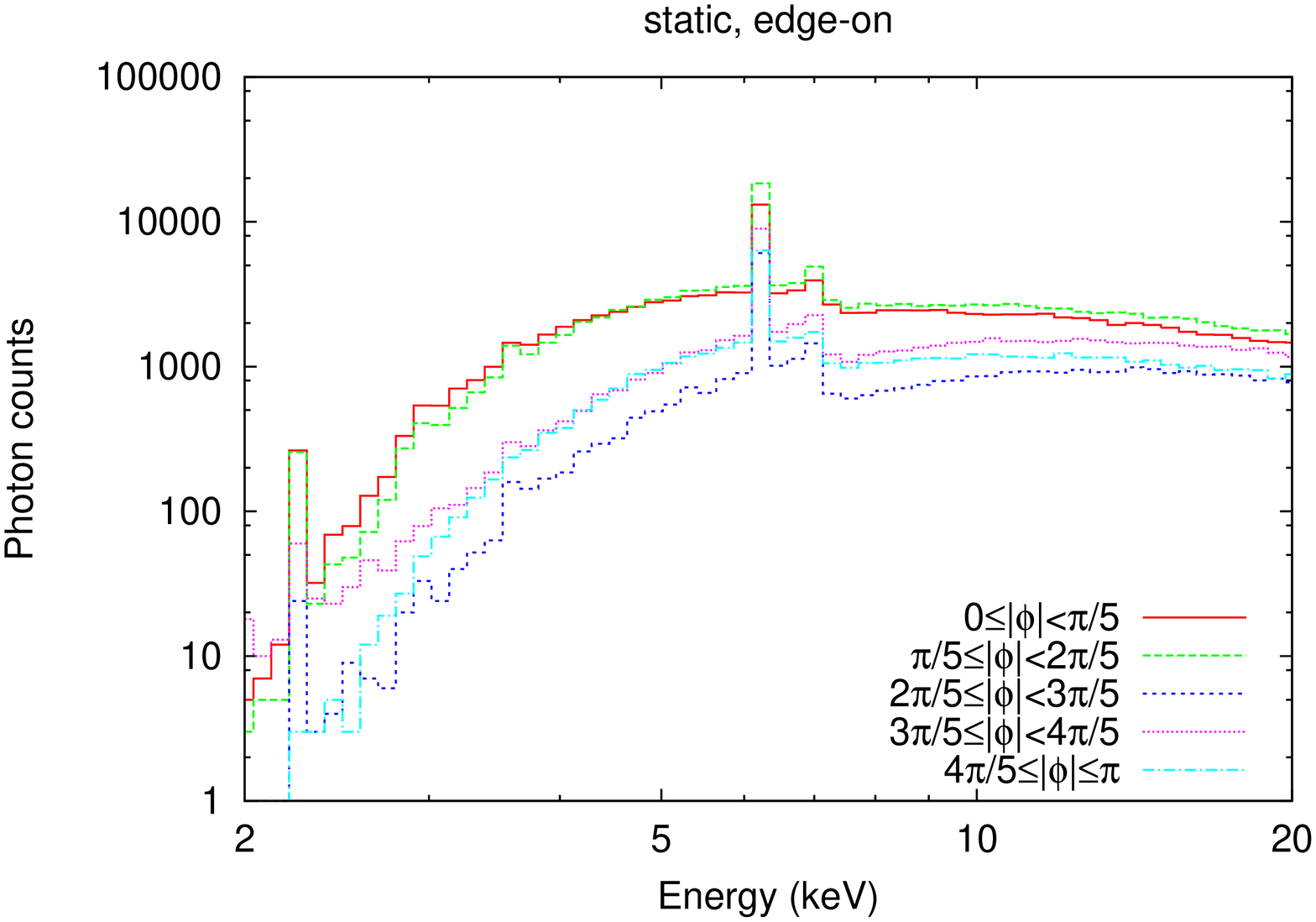}}
\resizebox{8cm}{!}{\includegraphics[angle=270]{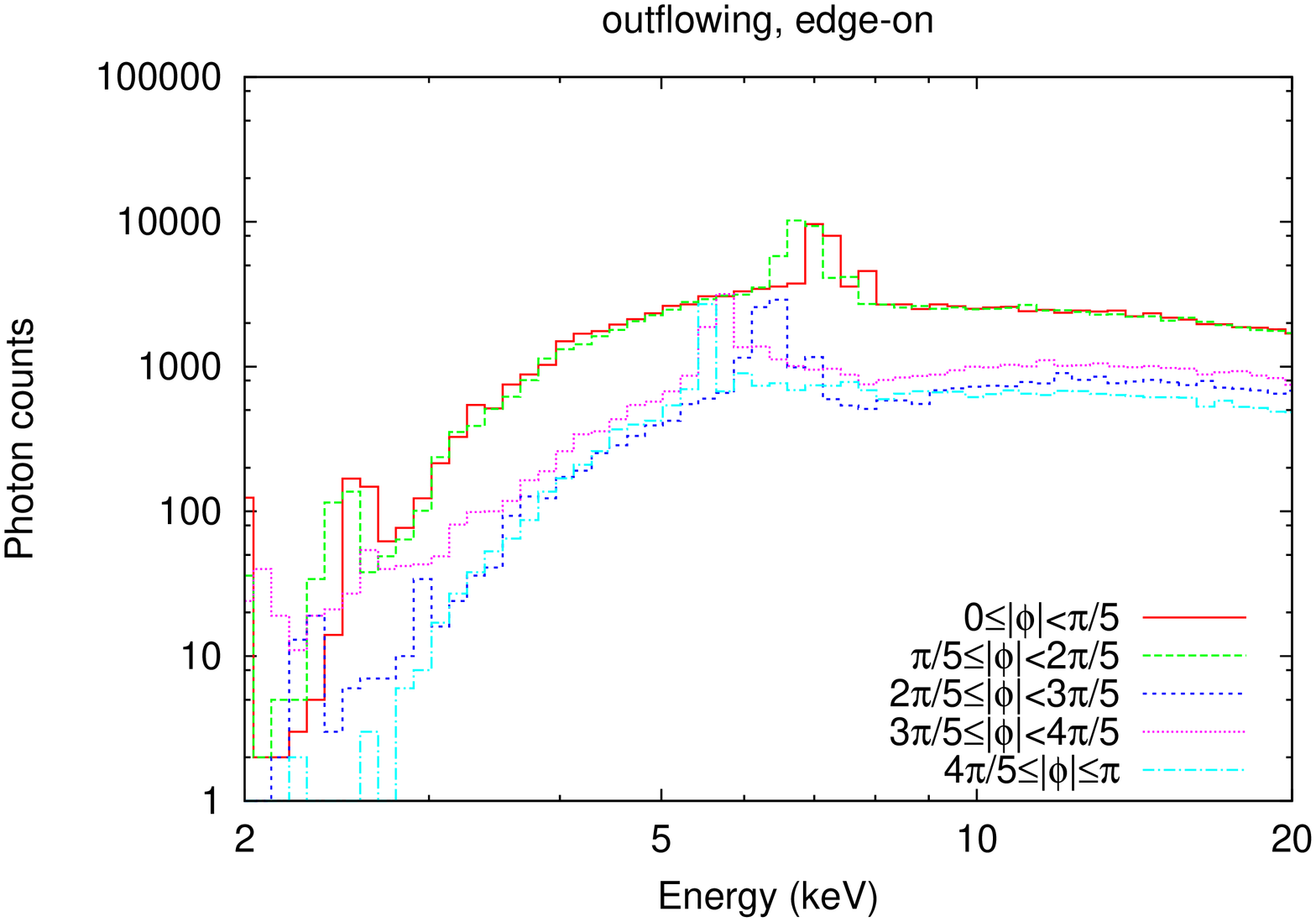}}
}
\subfigure{
\resizebox{8cm}{!}{\includegraphics[angle=270]{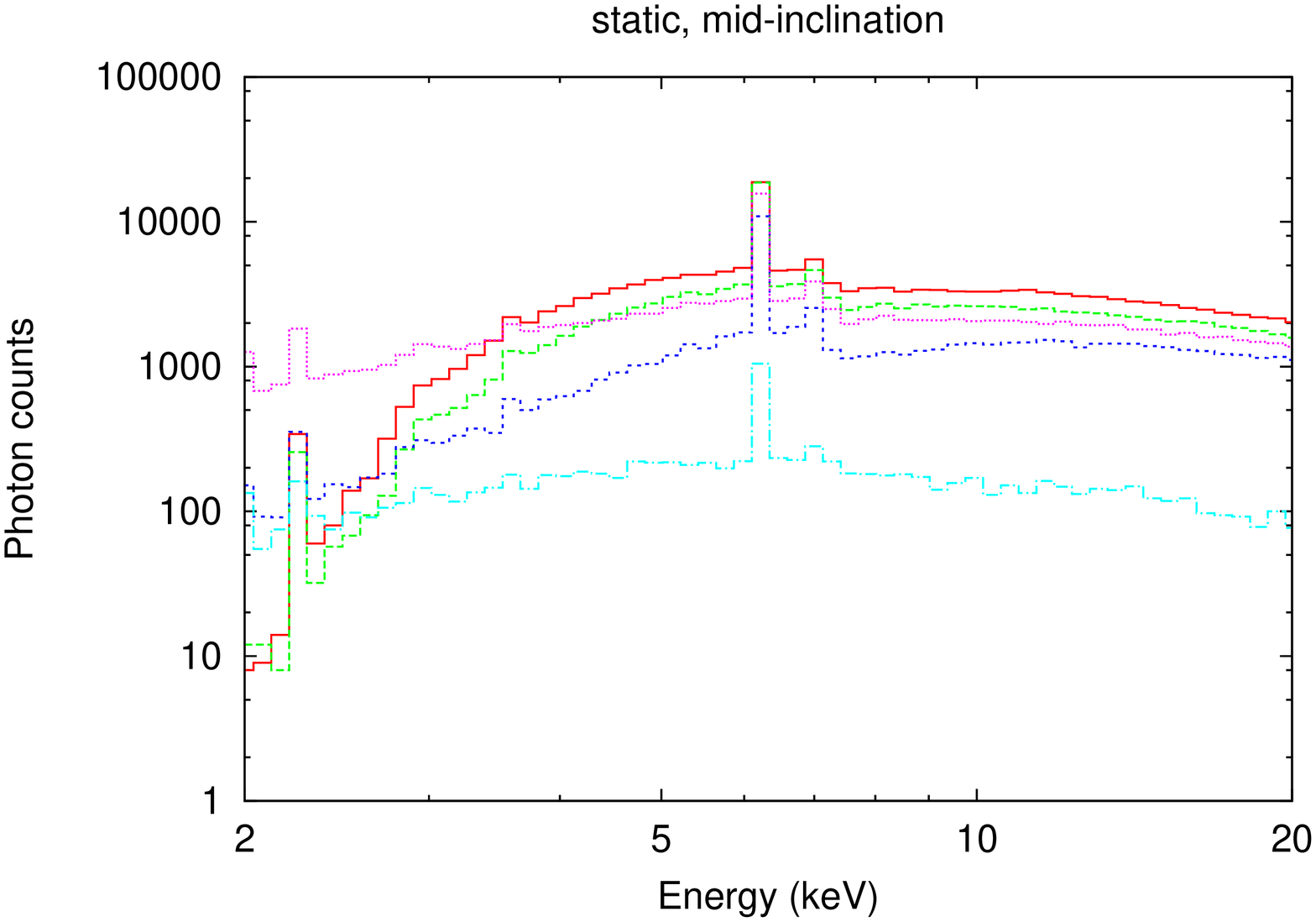}}
\resizebox{8cm}{!}{\includegraphics[angle=270]{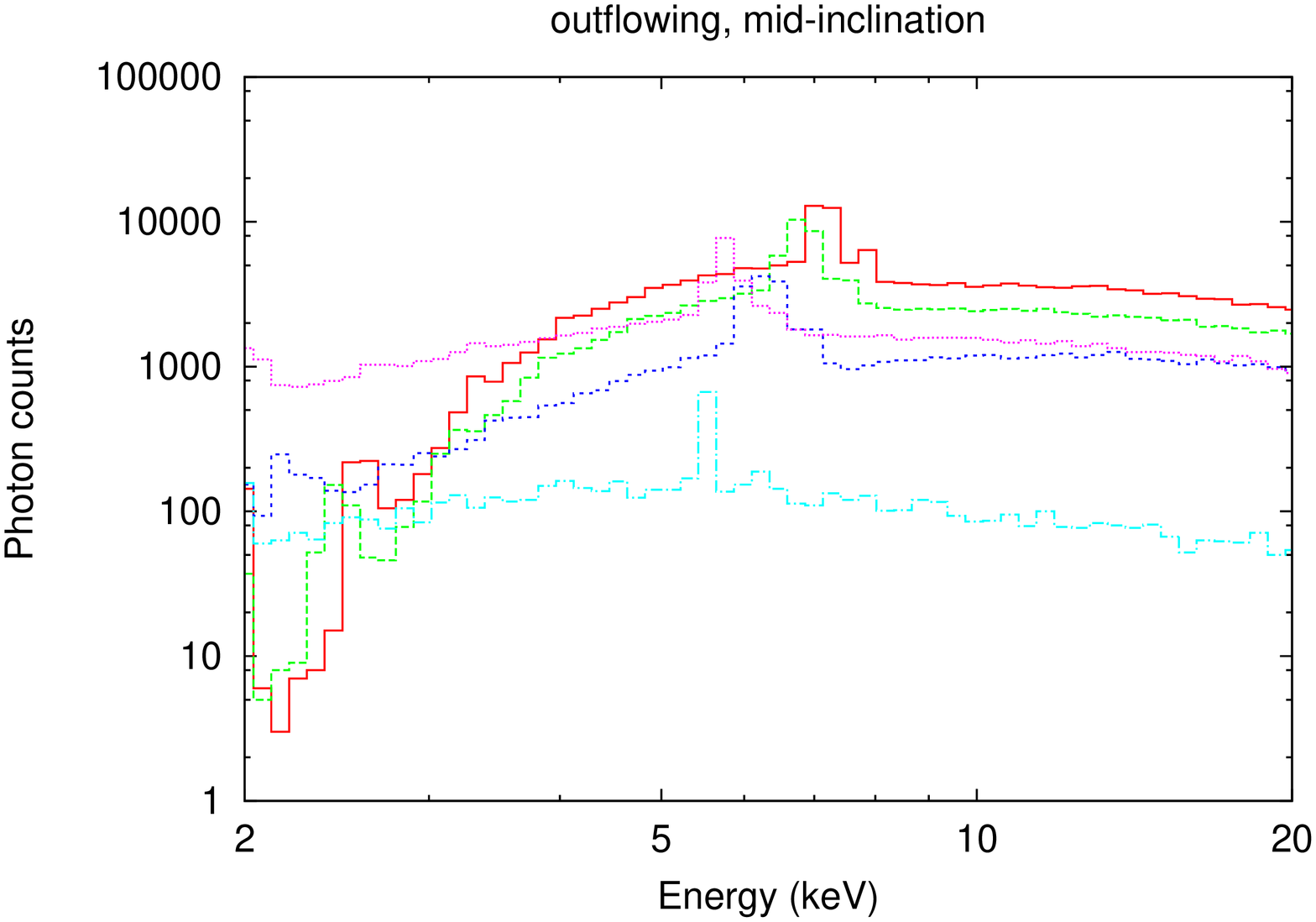}}
}
\subfigure{
\resizebox{8cm}{!}{\includegraphics[angle=270]{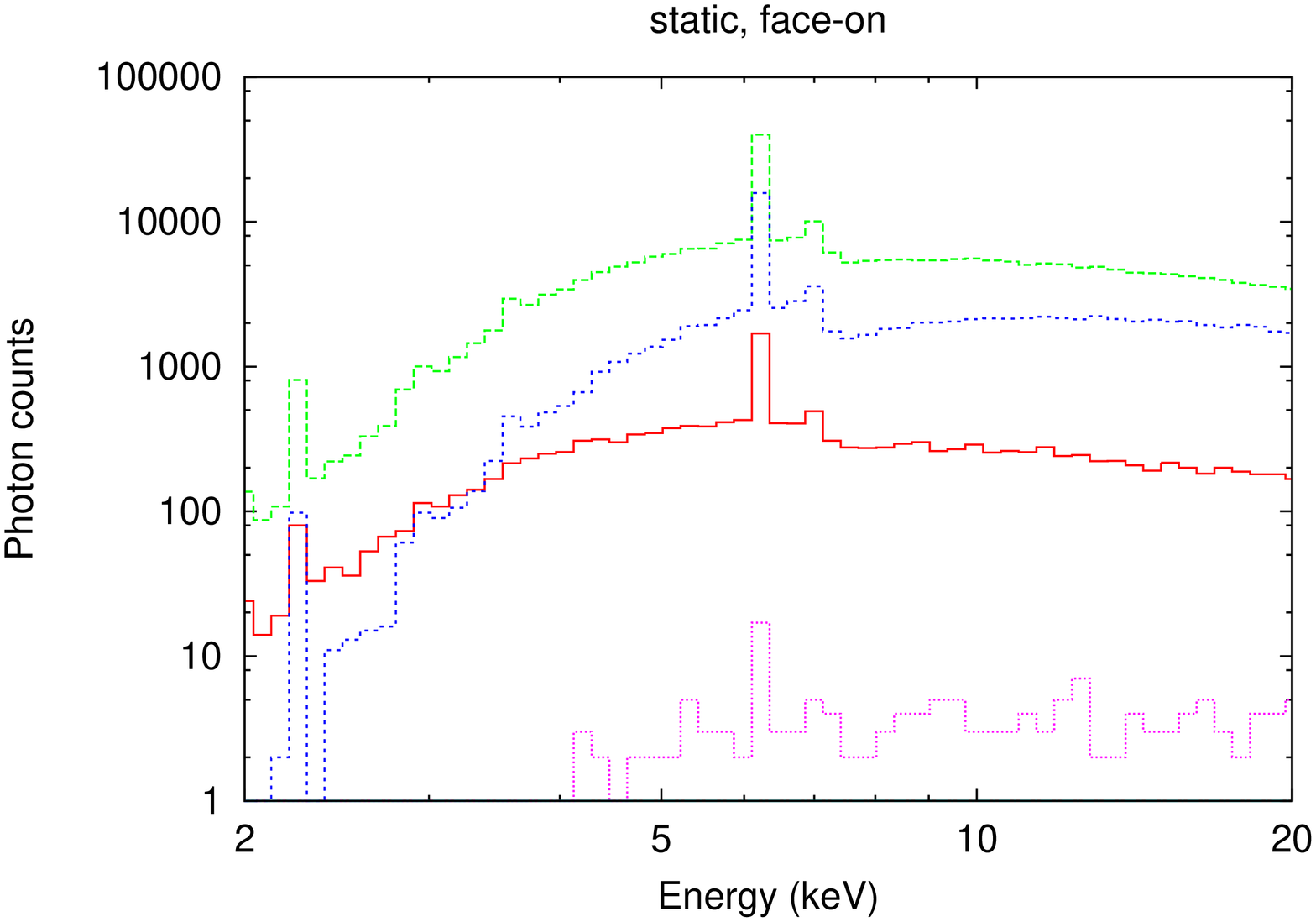}}
\resizebox{8cm}{!}{\includegraphics[angle=270]{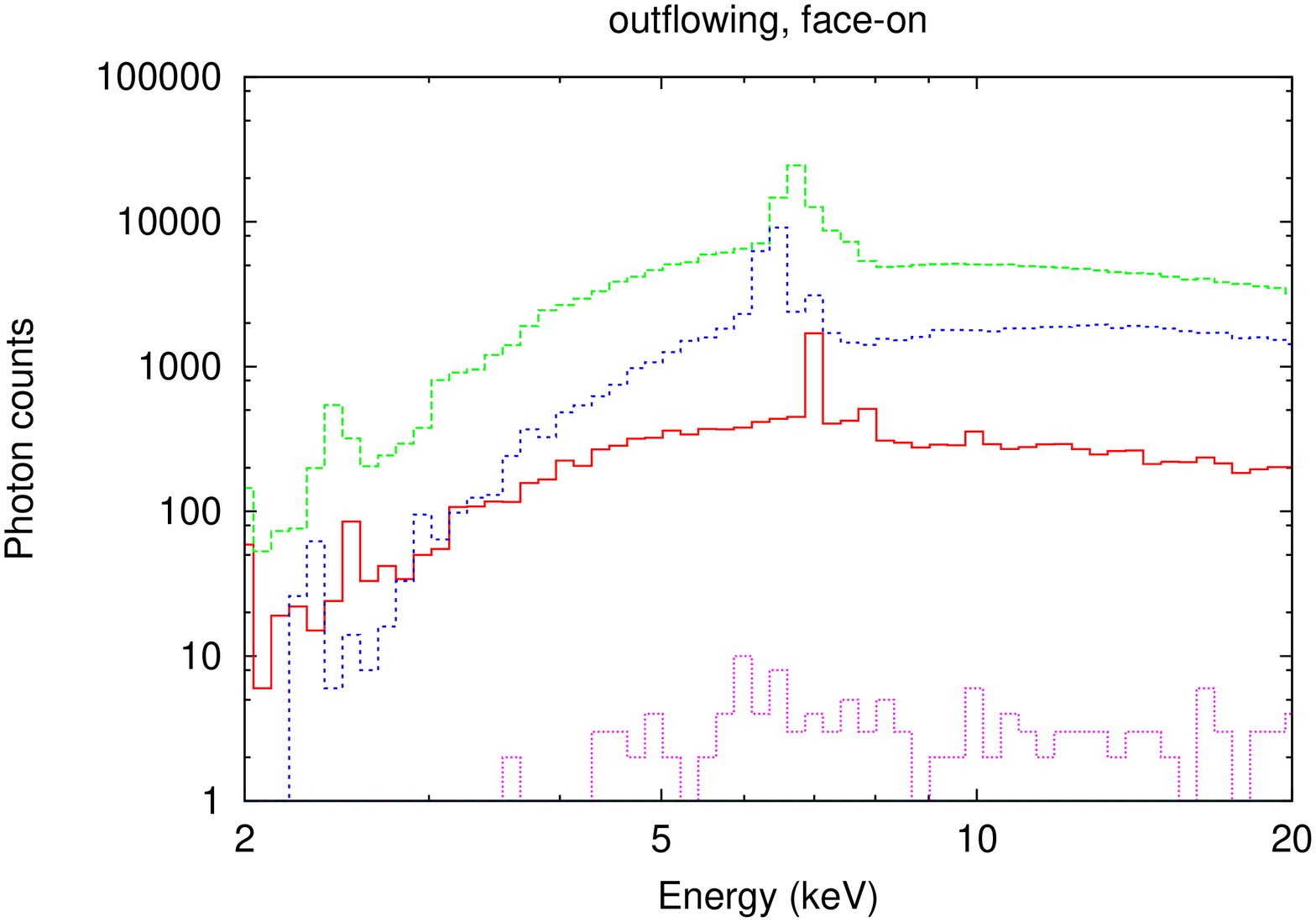}}
}
 \caption{Energy spectra of the reprocessed components with different scattering angles ($\phi$), for each of the static shell (left) and the outflowing shell (right). 
 The scattering angles are shown in the top panels.
 }\label{fig:anglespec}
 \end{center}
\end{figure*}

\section{Results} \label{sec3}

\subsection{Spectra and transfer functions}

\begin{figure*}
 \begin{center}
\subfigure{
\resizebox{8cm}{!}{\includegraphics[angle=270]{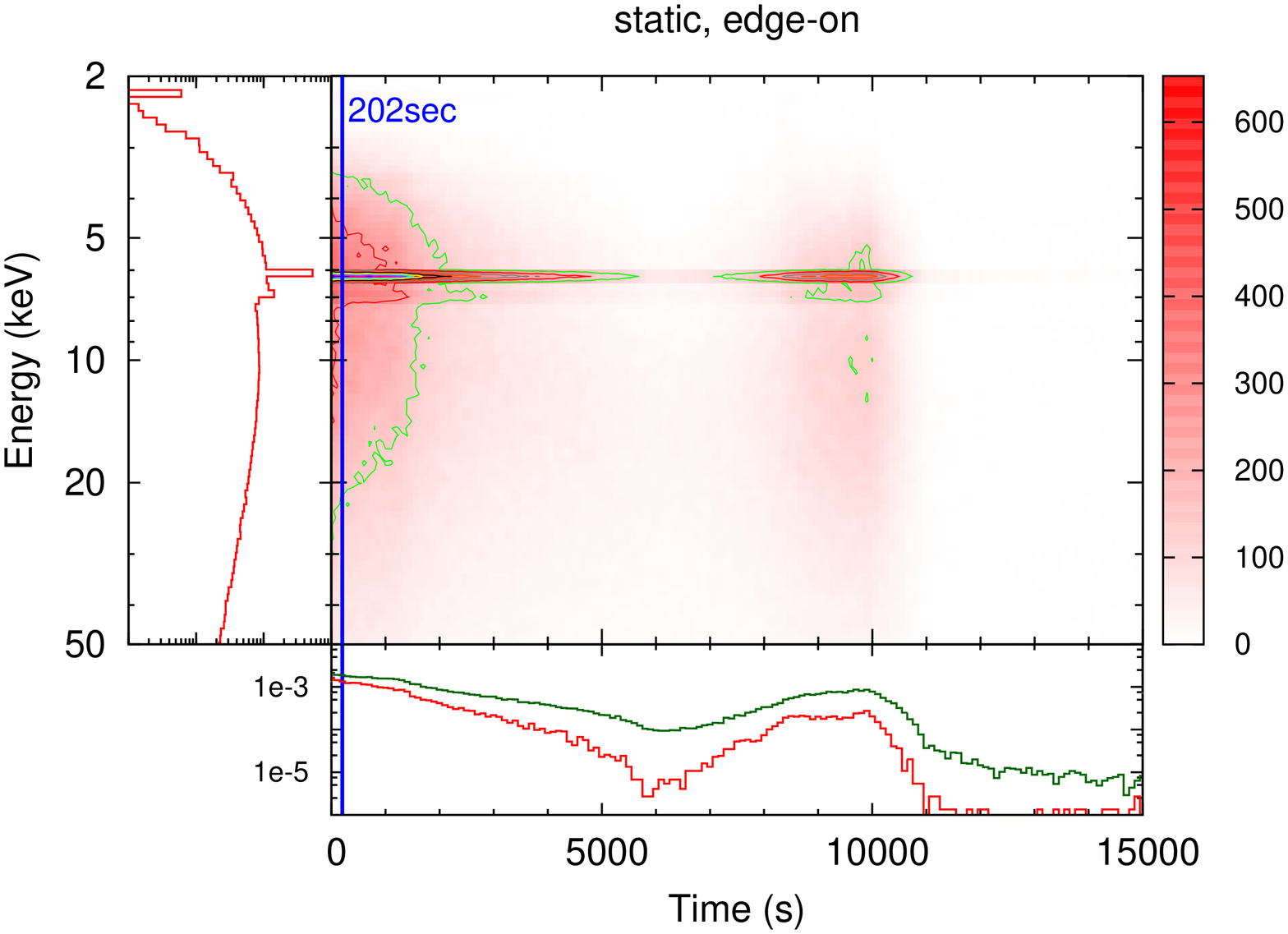}}
\resizebox{8cm}{!}{\includegraphics[angle=270]{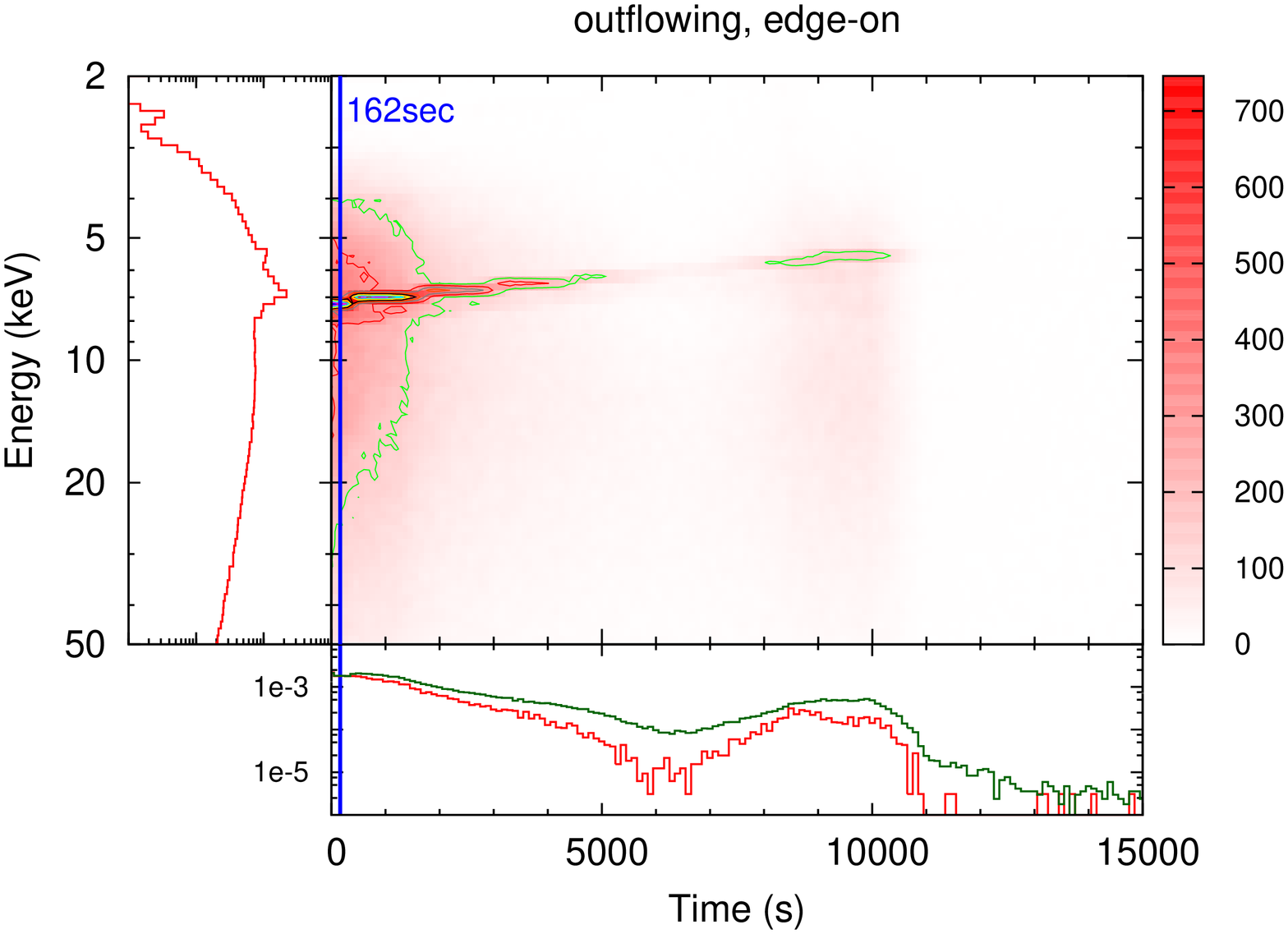}}
}
\subfigure{
\resizebox{8cm}{!}{\includegraphics[angle=270]{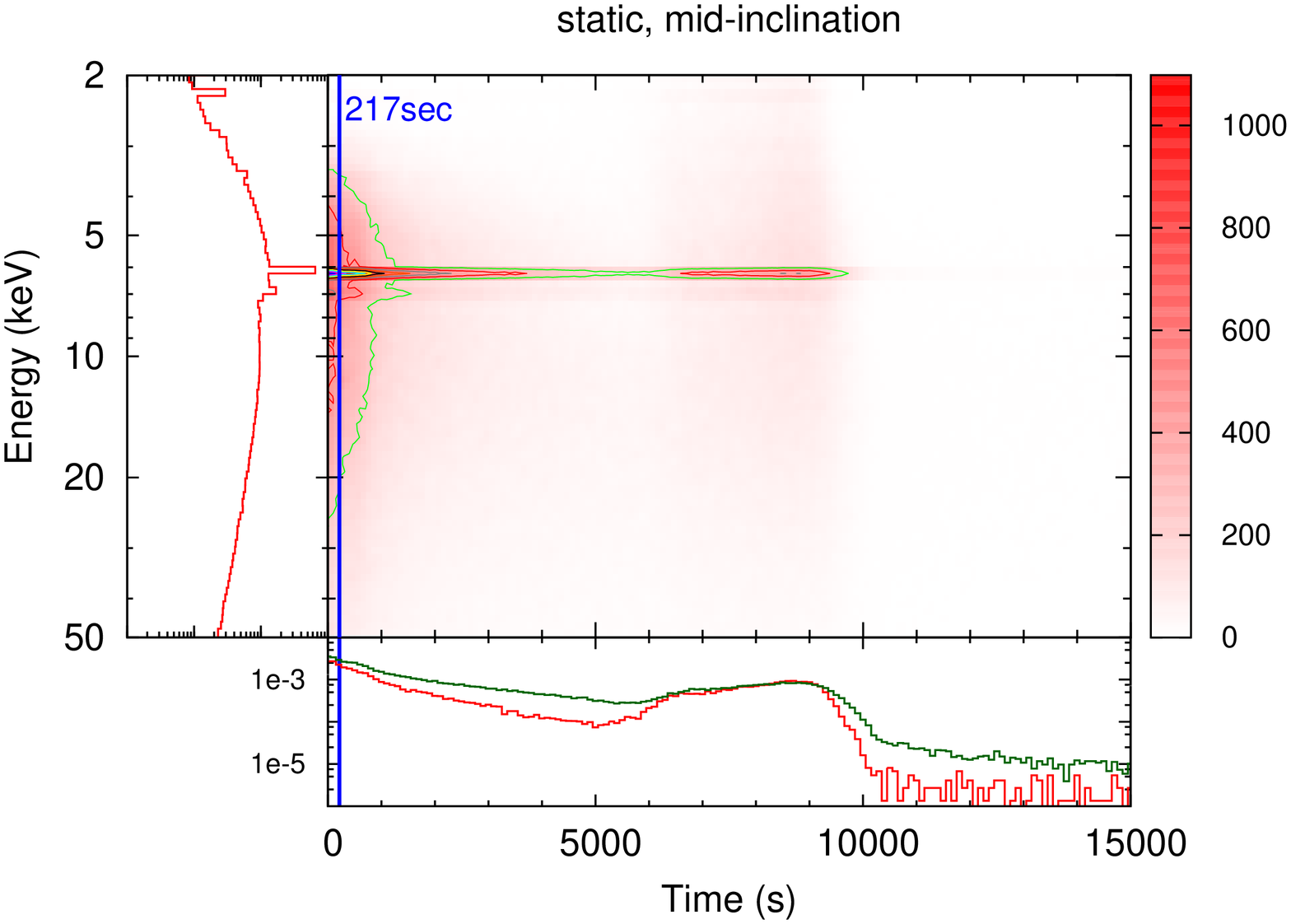}}
\resizebox{8cm}{!}{\includegraphics[angle=270]{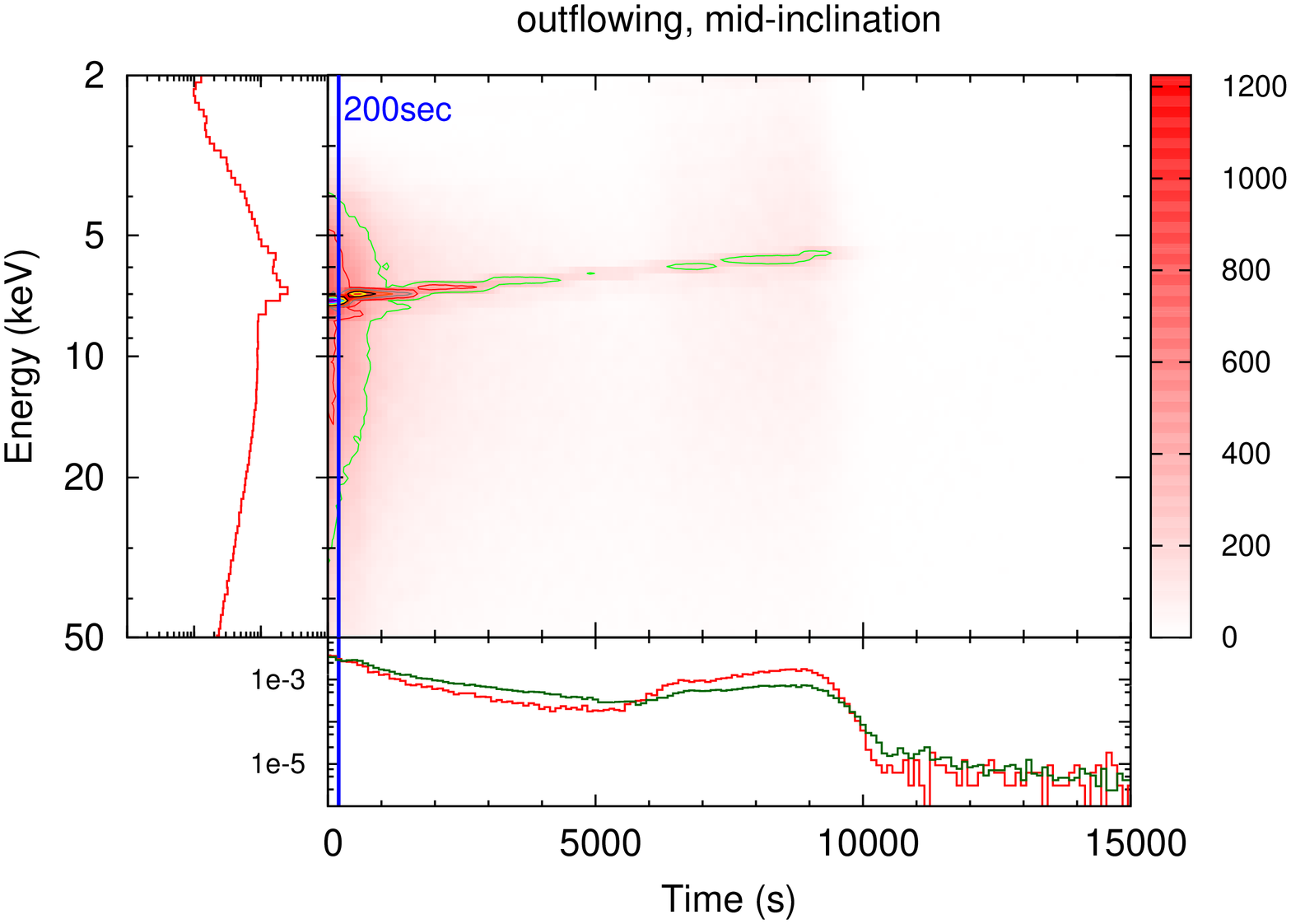}}
}
\subfigure{
\resizebox{8cm}{!}{\includegraphics[angle=270]{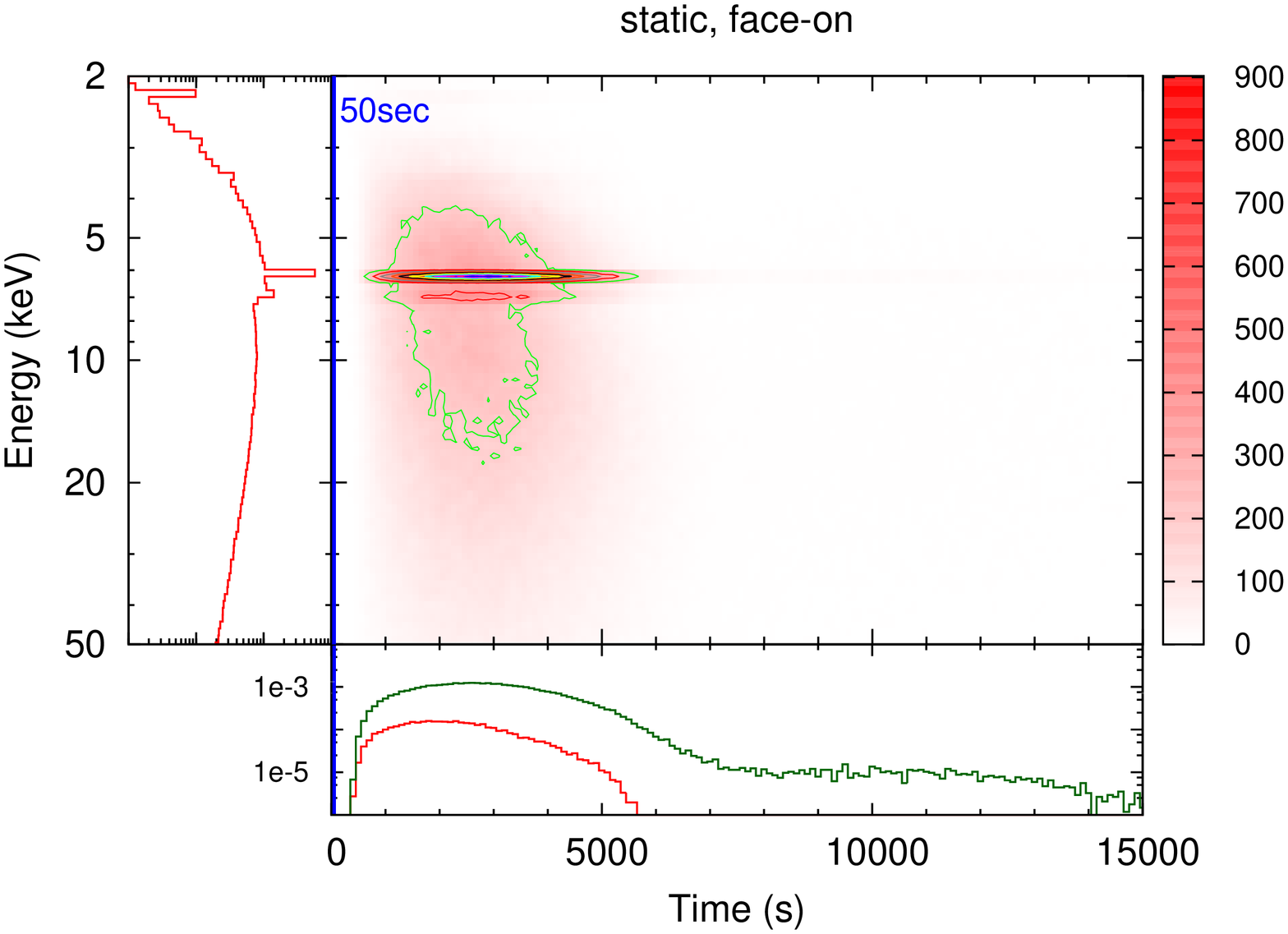}}
\resizebox{8cm}{!}{\includegraphics[angle=270]{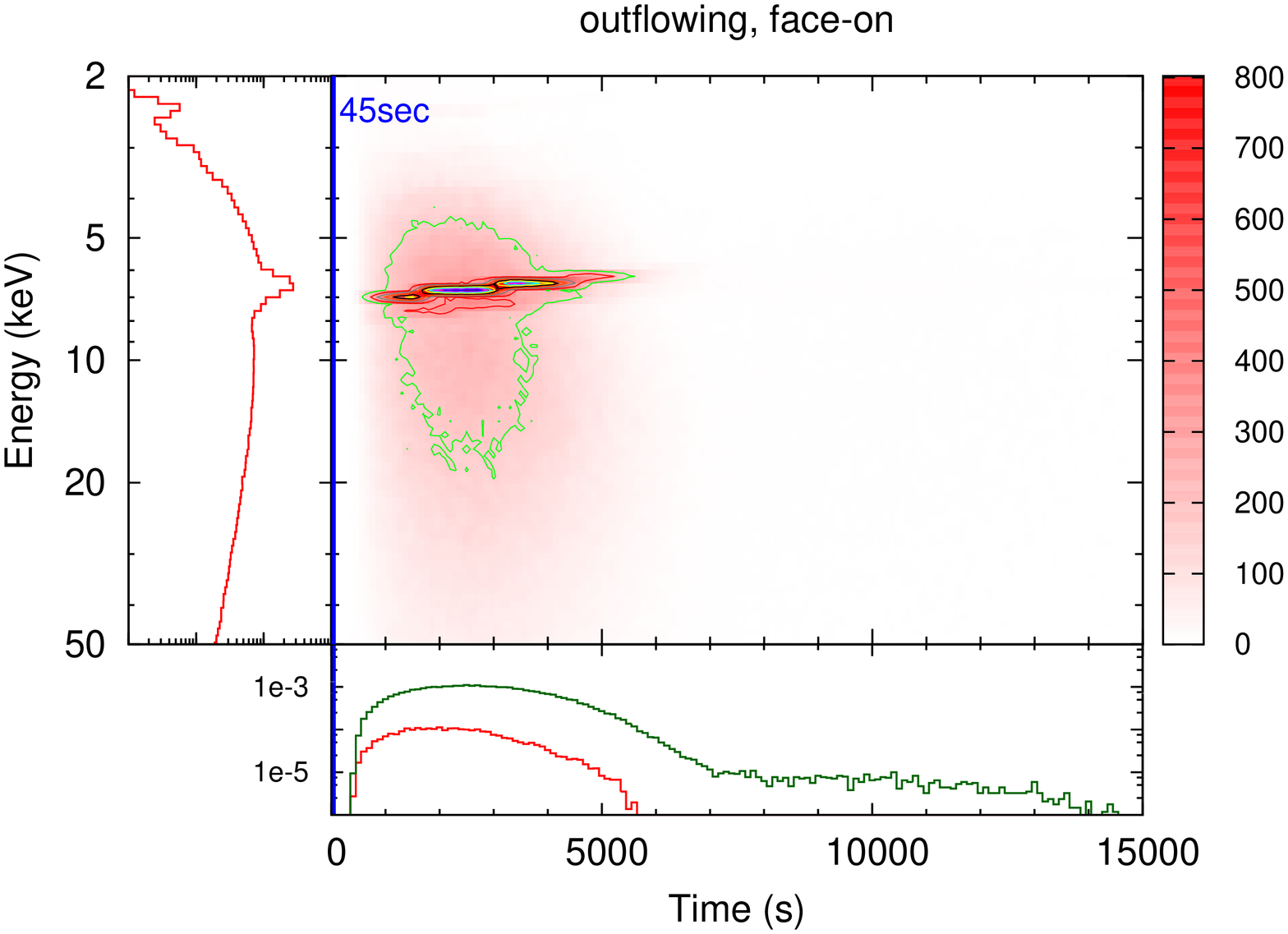}}
}
\caption{ Main panels: Two dimensional transfer function $R(E,t)$ for
  the static (left) and wind (right) geometries for a distant shell at
  $R=10^2\,R_g$ around the $M_\mathrm{BH}=10^7\,M_\odot$ black hole
  ($R/c=5000$~s), at different inclination angles. The green contour
  shows where the count is 0.1 of the peak count.  The blue vertical
  lines show the average delay time of all photons including primary ones. This is
  much smaller than $R/c$ as there are many more primary photons with zero
  lag than reprocessed photons with lag $\sim R/c$. The contour plots
  show where the response is at one-tenth of the
  peak count.  The left panel on each plot shows the two dimensional response
  integrated over time, giving the reprocessed
  spectra. The bottom panel on each plot shows the two dimensional response
  integrated over the 3-5~keV (red) and 5-7~keV (green) energy bands,
  normalised with respect to the primary component.The difference in
  response at these two energies gives the observed lags between these
  two bands.  }\label{fig:tf}
 \end{center}
\end{figure*}

The resultant spectra from the Monte-Carlo simulations have two
components; the primary one without time delay and the reprocessed one
with a range of time delays.  Fig.\ \ref{fig:spectrum} shows these
spectral components for the three inclination angles, for each of the
static shell and the outflowing shell.  The primary component (red
lines) in the edge-on and mid-inclination case are strongly absorbed
by the near side of the shell in the line of sight.  The absorption in
the soft energy band is stronger in the outflowing shell, because the
source spectrum is redshifted to lower energies in the wind rest frame
due to the Doppler shift.  The reprocessed components (green lines)
have fluorescent Fe-K lines at rest energies of 6.40~keV and 7.05~keV,
an Fe-K edge at 7.1~keV, and a Compton hump above $\sim$20~keV.  The
Fe-K features in the outflowing shell are broadened, whereas they are
sharp in the static shell.

Fig.\ \ref{fig:anglespec} shows the reprocessed spectra for the static
and outflowing cases for different scattering angles ($\phi$).  This
is mainly determined by the azimuthal angle of the point of last
scattering, so this also sets the light-travel time lag.  Photons with
the longest lag time 
come from the far side of the wind so these are
redshifted, whereas those with the shortest lag time are from the near
side of the wind, so are blueshifted, while photons from the sides,
with light travel time $\sim R/c$ are at the rest energy (though this
will change if rotation of the wind is included). 
The superposition of 
these different Fe-K energies gives a broad 
line feature in the integrated response (Fig.\
\ref{fig:spectrum}, right panels). 

In the edge-on case, photons in the soft energy bands
are absorbed in all $\phi$ values.  In the mid-inclination case,
soft photons are not absorbed when $\phi$ gets larger as the far
side of the shell can be seen directly even when the central source is
obscured. In the face-on case,
there are few photons whose $\phi$ is larger than $3\pi/5$.  
When an incident photon is emitted in the $x$ direction,
$\vec{p}=(-1,0,0)$, and $\vec{q}=(\sqrt{29}/15,0,14/15)$, the maximum 
$\phi$ value is 111$^\circ$.04, only slightly larger than
$3\pi/5\,\mathrm{rad}=108^{\circ}$.

Fig.\ \ref{fig:tf} shows counts of reprocessed photons in each time-
(bin size of $100$~s) and energy-bin (bin size of $\Delta\log E\,({\rm
  keV})=0.017$) after the delta function continuum injection at $t=0$.
This corresponds to the ``two-dimensional transfer function'' times
the input photon spectrum.  The left panels show projection on the 
$y$-axis, which gives the total energy spectrum of all the reprocessed
photons.  This picks out the energy bands of iron emission lines and
Compton hump where the reprocessed photons are concentrated.
The bottom panels plot counts of the reprocessed
  photons in 3--4~keV (red) and 5--7~keV (green) energy bands against
  the delay time. This corresponds to the response function in these
  bands (see e.g.\ \citealt{epi16}). The measured lag between these
  bands is the   difference between these two responses. 

It is easy to see that the normalisation of the response is higher
  in the 5--7~keV band than in the 3--4~keV band for all simulations
  except for the outflowing, mid inclination case discussed
  previously. Thus the hard band will lag behind the soft as it has a
  higher contribution from the reprocessed flux.  In both the edge-on
  and the mid-inclination cases, the response functions have two
  peaks, the first from 0--4000~s ($\lesssim1R/c$) from scattering on
  the near side of the shell with a small scattering angles $\phi$,
  and the second peak at $\sim9000$~s ($\simeq2R/c$) from scattering
  on the far side of the shell with large $\phi$.  These two merge
  into a single peak for the face-on case. These all span the range in
  lag times predicted in Section \ref{sec2}, i.e., $\sim 0-2R/c$ for edge on
  inclination and $\sim (1-\sin\theta_{\rm op})R/c-R/c$ for face on.

The blue vertical lines show the average delay time, i.e., the sum of
all the reprocessed photons at all energies times their lag time,
divided by the total number of photons (primary plus reprocessed over
the entire energy band).  The number of reprocessed photons is much
less than that of the primary photons, and thus the average delay time
is much shorter than the light-travel time of the reprocessed photons
({\it dilution effect}).  In the edge-on and mid-inclination cases,
the average delay time is $\sim$200~s, corresponding to $4\,R_g/c$.
This is much shorter than the light-travel time from the source to the
reflection shell by more than one order of magnitude.  In the face-on
case, the average delay time is even shorter, $\sim$50~s, which
corresponds to $R_g/c$, because there are more primary photons and
thus the dilution effect is more significant.

\subsection{Lag-frequency plot}\label{sec:lagf}

The frequency-dependent lag, $\tau(f)$, is defined as
\begin{equation}
\begin{split}
\tau(f)&=\mathrm{arg}[C(f)]/(2\pi f) \\
C(f)&= \mathcal{S}(f)\mathcal{H}^*(f)\label{eq1},
\end{split}
\end{equation}
where $^*$ denotes complex conjugate and $\mathcal{S}(f)$ and $\mathcal{H}(f)$ are 
Fourier transforms of soft- and hard-band light curves, $s(t)$ and $h(t)$ \citep{vau97,now99}, which is called as a cross spectrum.
We use the standard convention that a positive lag means that photons in the hard band lag behind those in the soft band (e.g., \citealt{kar13b}).
In other words, $s(t)$ is a light curve in a reference band, and $h(t)$ is the one in an energy band of interest.

An observed light curve at a given energy bin $E$ is expressed as 
the sum of the primary emission and the reprocessed emission.
The primary emission is expressed as $P(E)g(t)$,
where $P(E)$ is the (power-law) spectrum of the primary component and
$g(t)$ is the intrinsic flux variability, 
which is assumed to have no $E$-dependence, i.e., 
the continuum varies only in normalisation. 
We define $R(E,k)$ as photon counts of the reprocessed component in each energy-bin ($E$) and time-bin ($k$),
which is shown in Fig.\,\ref{fig:tf}.
The reprocessed emission is expressed
as $\sum_{k=0}^{k_{\rm max}} R(E,k) g(t - k t_{\rm bin})$,
where $t_\mathrm{bin}$ is a time bin-size of the light curve ($=100$~s in this paper).
We take $k_{\rm max}=400$, so we consider lags between 0--40,000~s. 
Here, the light curve at a given energy bin $E$ is written as
\begin{equation}
l(E,t)=P(E) g(t) + \sum_{k=0}^{k_{\rm max}} R(E,k) g(t- k t_{\rm bin}) \label{eq:l},
\end{equation}
and its Fourier transform is expressed as
\begin{equation}
\mathcal{L}(E,f)=P(E) \mathcal{G}(f) + 
\sum_{k=0}^{k_{\rm max}} R(E,k) \exp[-2\pi i(kt_\mathrm{bin})f]\mathcal{G}(f)\label{eq:lf},
\end{equation}
where $\mathcal{L}(E,f)$ and $\mathcal{G}(f)$ are Fourier transforms of $l(E,t)$ and $g(t)$.
Thus the soft- and hard-band light curves and their Fourier transforms are expressed as
\begin{equation}
\begin{split}
s(t)&=\sum_{E\in{\rm \,soft\,band}}l(E,t) \\
h(t)&=\sum_{E\in{\rm \,hard\,band}}l(E,t), \label{eq:ltcrv}
\end{split}
\end{equation}
and 
\begin{equation}
\begin{split}
\mathcal{S}(f)&=P_s \mathcal{G}(f) + \sum_k R_s(k)\exp[-2\pi i(kt_\mathrm{bin})f]\mathcal{G}(f) \\
\mathcal{H}(f)&=P_h \mathcal{G}(f) + \sum_k R_h(k)\exp[-2\pi i(kt_\mathrm{bin})f]\mathcal{G}(f), \label{eq:Fourier}
\end{split}
\end{equation}
where $P_s=\sum_{E\in{\rm \,soft\,band}}P(E)$, $R_s(k)=\sum_{E\in{\rm \,soft\,band}}R(E,k)$, and
$P_h$ and $R_h(E)$ are those in the hard band.
Therefore, equation (\ref{eq1}) is calculated as
\begin{align}
C(f)&=\mathcal{S}(f)\mathcal{H}^*(f) \nonumber\\
&= \left(P_s+\sum_k R_s(k)\exp[-2\pi i(kt_\mathrm{bin})f]\right) \nonumber \\
& \:\:\:\:\:\:\left(P_h+\sum_k R_{h}(k)\exp[2\pi i(kt_\mathrm{bin})f]\right)|\mathcal{G}(f)|^2 \label{eq:Ccalc}
\end{align}
and
\begin{align}
\tau(f)&=\frac{1}{2\pi f}\;\mathrm{arg}\left[\left(P_s+\sum_k R_s(k)\exp[-2\pi i(kt_\mathrm{bin})f]\right)\right.\nonumber\\
& \:\:\:\:\:\:\:\:\:\:\:\:\:\:\:\:\:\:\:\:\:\:\:
\left.\left(P_h+\sum_k R_h(k)\exp[2\pi i(kt_\mathrm{bin})f]\right)\right]. \label{eq:lagdef}
\end{align}
Note that the frequency-dependent lag amplitudes do not depend on the
functional form of the intrinsic variation,
as long as the lags are caused by the reverberation.
Hereafter, we calculate lags based on equation (\ref{eq:lagdef}).

\begin{figure*}
 \begin{center}
\subfigure{
\resizebox{8cm}{!}{\includegraphics[angle=270]{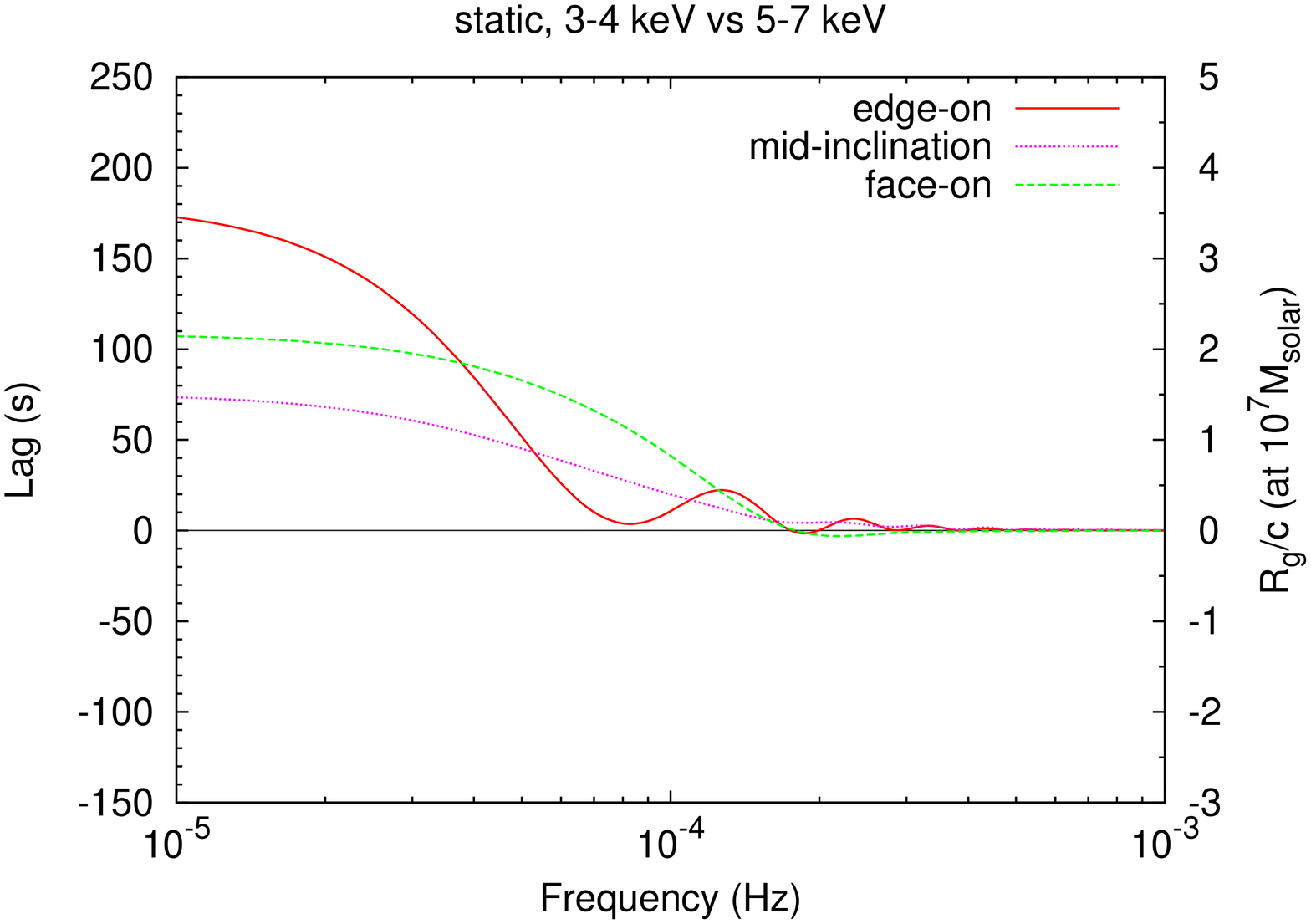}}
\resizebox{8cm}{!}{\includegraphics[angle=270]{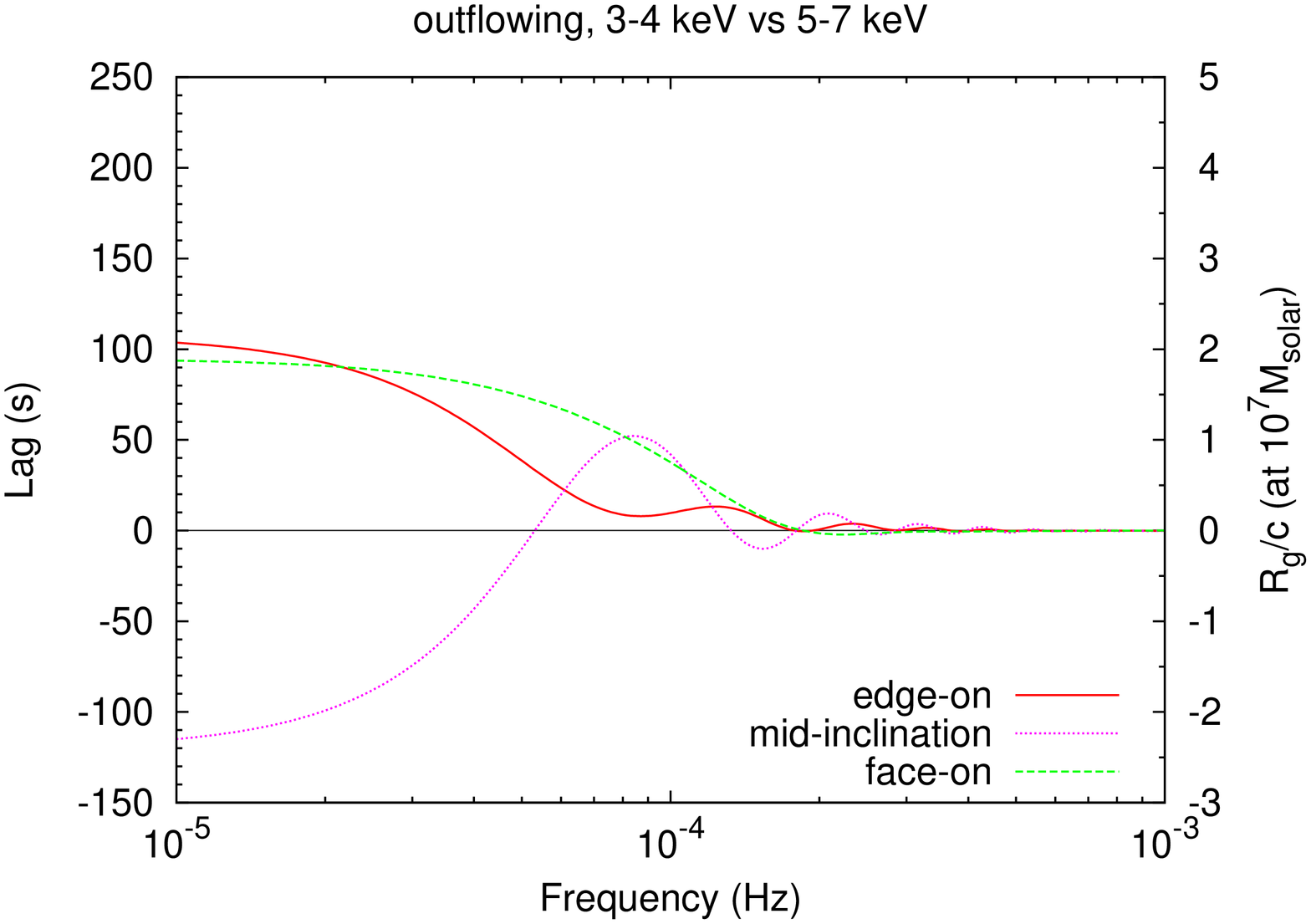}}
}
 \caption{Lag-frequency plots when comparing 3-4 keV with 5-7 keV, derived from the Monte-Carlo simulation.
Positive means hard lags, and vice versa.
 }\label{fig:lagvsf}
 \end{center}
\end{figure*}

Fig.\ \ref{fig:lagvsf} shows the lag-frequency plots, comparing a soft
band (3--4 keV, where there is (generally) least contribution from the
reprocessed component so that this is the best estimator of the
continuum light curve; Fig.\ \ref{fig:spectrum}) with a hard band
(5--7 keV, where the Fe-K emission peaks).  In all cases except for
the outflowing mid-inclination case, {\it hard} lags are seen with an
amplitude of $\lesssim150$~s, which is much shorter than the
light-travel time to the shell (5000~s) and comparable to the weighted
average delay time (Fig.\ \ref{fig:tf}).  In the outflowing
mid-inclination case, {\it soft} lags with an amplitude of $\sim100$~s
are seen because the reference 3--4~keV band has a higher fraction of
reprocessed flux in this case than the others, so the ``continuum''
band is actually more dominated by lagged photons than the other
cases.  The lags are attenuated in all the cases for frequencies
$>2\times10^{-4}$~Hz, because variations on timescales faster than
$R/c$ are smeared
by the light-travel time across the shell.

We can get a useful analytic expression for the dilution by
approximating the full response function by a single (average) intrinsic delay
time, $\Delta \tau$($=k_0t_{\rm bin}$). Then 
equation (\ref{eq:lagdef}) can be evaluated in a simple way as
equation (\ref{eq:ltcrv}) becomes
\begin{equation}
\begin{split}
s(t)&=P_{s}g(t)+R_{s}g(t-\Delta\tau) \\
h(t)&=P_{h}g(t)+R_{h}g(t-\Delta\tau), \label{eq:singletau}
\end{split}
\end{equation}
where $R_s$ and $R_h$ are the reprocessed component in the soft- and hard-band. 
In this case, $\tau(f)$ is calculated as
\begin{align}
\tau(f)
&= \frac{1}{2\pi f}\mathrm{arg}\left[\left(P_s+R_{s}\exp[-2\pi i\Delta\tau f]\right)
\left(P_h+R_h\exp[2\pi i\Delta\tau f]\right)\right] \nonumber\\
&= \frac{1}{2\pi f}\arctan\left(\frac{(P_sR_h-R_sP_h)\sin(2\pi \Delta\tau f)}{P_sP_h+R_sR_h+(P_sR_h+R_sP_h)\cos(2\pi\Delta\tau f)}\right). \label{eq:dilution}
\end{align}
In the $f\rightarrow0$ limit,
\begin{equation}
\tau(f\rightarrow0)
=\frac{P_sR_h-R_sP_h}{(P_s+R_s)(P_h+R_h)}\Delta\tau
=\mathrm{DF}\Delta\tau,  \label{eq:dilution2}
\end{equation}
where the dilution factor DF can be calculated directly
from the primary and reprocessed emission in each band.
For example, when we use the parameters of the static edge-on case,
then the soft band (3--4~keV) counts in the primary and reprocessed flux give 
$P_s=34.1$ and $R_s=1$, while the hard band (5--7~keV) has $P_h=111.6$ and
$R_h=7.2$ (see Fig.\ \ref{fig:spectrum}), so ${\rm DF}=3.17\times 10^{-2}$. 
This predicts that DF reduces the observed lags by more than one order of magnitude. 
The mean intrinsic lag is of order $\Delta \tau\sim R/c=5000$~s, so 
this predicts an observed lag at low frequencies of $158$~s, 
which is very close to the $\sim 170$~s seen in the right panel of Fig.\ \ref{fig:lagvsf} for the edge-on case (red-solid line) as $f\to 0$.
The difference comes from the mean intrinsic lag time being different
from $R/c$ by a small factor
(see the energy-compressed response in Fig.\ \ref{fig:tf}).

\begin{figure}
  \begin{center}
\includegraphics[angle=270,width=\columnwidth]{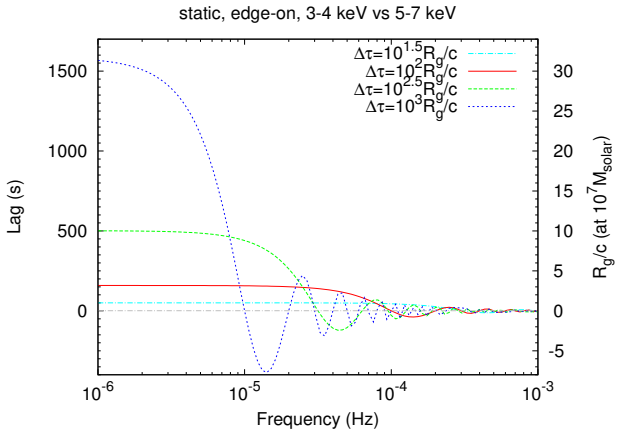}
  \caption{Lag-frequency plots with different size scales for the
    reprocessor assuming that the response is a 
delta-function at $R/c$. 
Parameters of the static edge-on case are used.
The time delay $\Delta\tau$ gets shorter from top to bottom at $10^{-6}$~Hz.
  }\label{fig:model}
  \end{center}
\end{figure}

Equation (\ref{eq:dilution2}) explicitly shows that
  the measured lag amplitudes are not a direct diagnostic of the size
  scale of the region; indeed they depend on the relative contribution
  of primary and reprocessed emission in each band.  However, the
  frequency at which the lags go to zero is a much more robust
  constraint. For the delta-function approximation, equation
  (\ref{eq:dilution}) shows that this occurs where $2\pi \Delta t f=
  n\pi$ where $n$ is an integer, so the first zero crossing picks out
  $f_0=1/(2\Delta t)=c/2R$~Hz. 
  Fig.\ \ref{fig:model} illustrates the light-travel time dependence of the
  lag frequencies, which is analytically calculated from equation
  (\ref{eq:dilution}) for different size scales of the
  reprocessor. The lags drop to zero at $f_0$, giving a clear
  diagnostic of the intrinsic size scale, and then oscillate
  around zero with  decreasing amplitude (see also Fig.\ 21 of \citealt{utt14}). 
 The observed lag frequencies are $\sim10^{-4}\,(10^7M_\odot/M_{\rm BH})$~Hz,
which corresponds to $R=100\,R_g$. This means that materials located at $R\lesssim100\,R_g$ contribute to the observed lag features.

However, the real response function 
covers a range in timescales from $\tau_{\rm min}<t<\tau_{\rm max}$  (see Fig.
\ref{fig:tf}) rather than being a delta function at a single response time. We repeat the analysis
assuming uniform response (a top-hat function: Appendix
  \ref{app1}). The spectral dilution factor is unchanged, so the lag as $f\to 0 $
is the same as before (see also Fig.\ 21 of \citealt{utt14}). 
However, the finite width of the response
  smooths out the oscillations around zero at high frequencies, and
  increases the attenuation so that the response drops more gradually.
  In the limit of a top hat from $0-2R/c$, the response never becomes
  negative, so there is no zero crossing, but there is still a minimum close
  to $\sim f_0$ (as shown in Fig.\ \ref{fig:lagvsf} where the
  relevant response shown in Fig.\ \ref{fig:tf} is approximately a top hat from
  $0-2R/c$). 

\subsection{Lag-energy plot}\label{sec:lagE}

\begin{figure*}
 \begin{center}
\subfigure{
\resizebox{8.5cm}{!}{\includegraphics[angle=270]{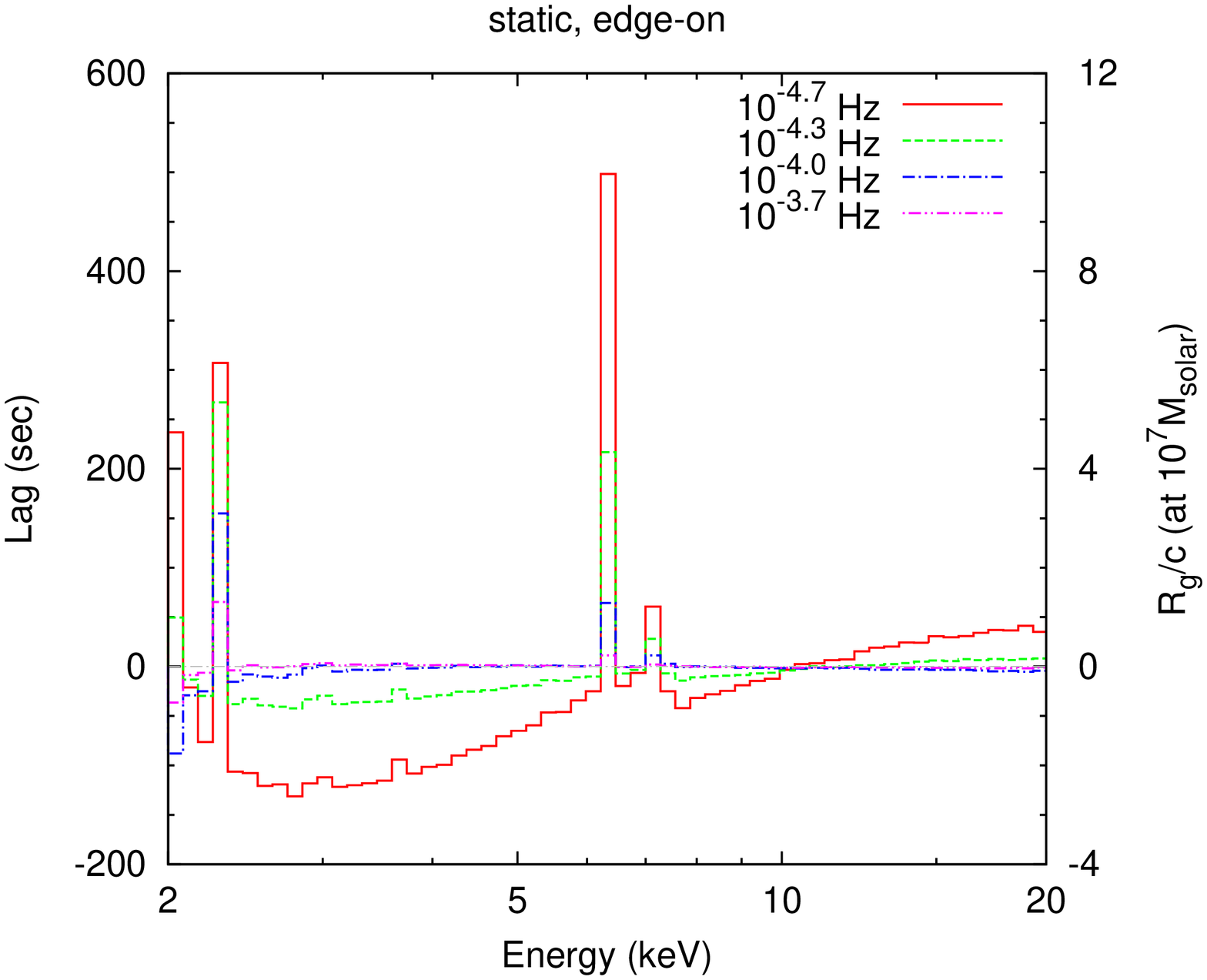}}
\resizebox{8.5cm}{!}{\includegraphics[angle=270]{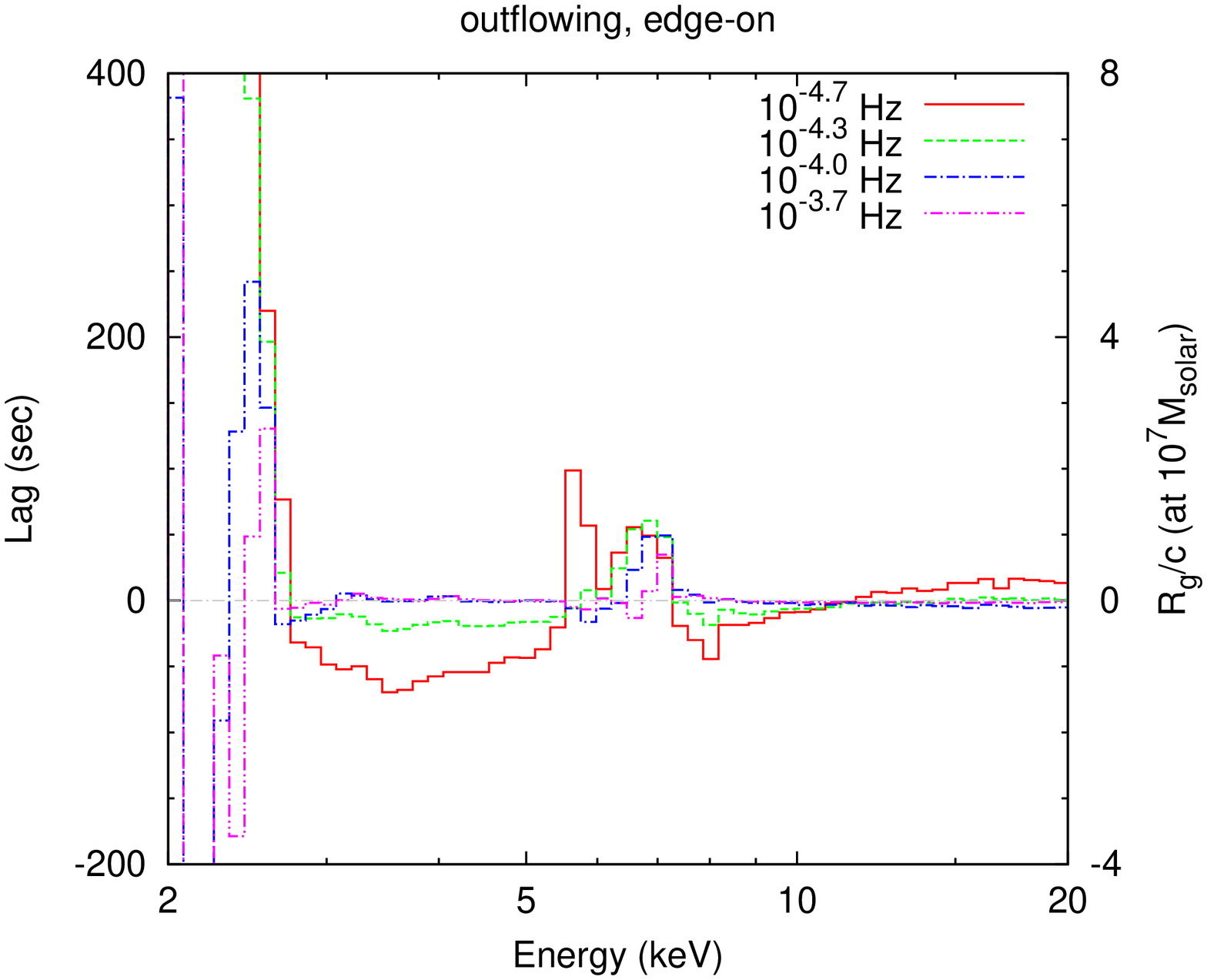}}
}
\subfigure{
\resizebox{8.5cm}{!}{\includegraphics[angle=270]{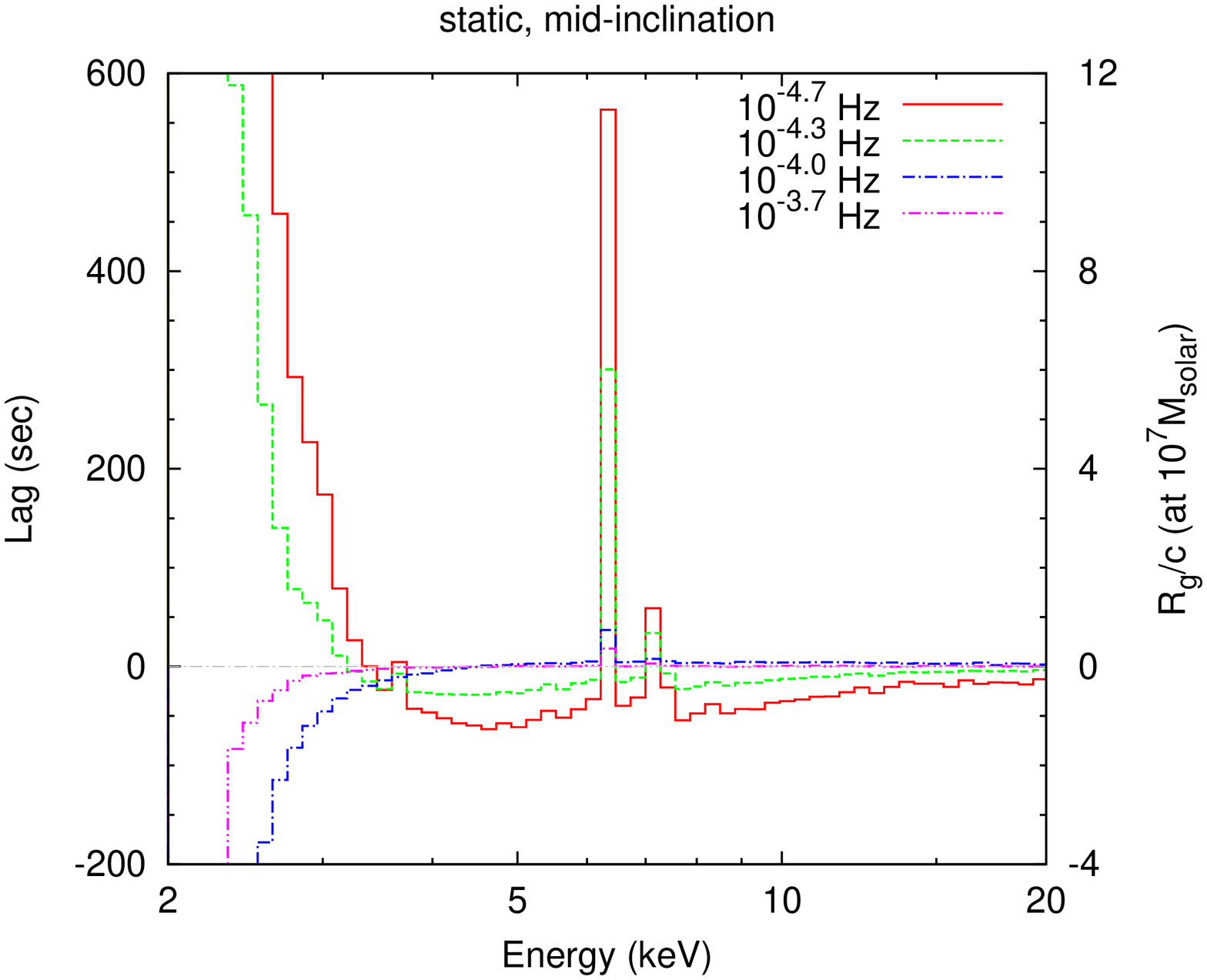}}
\resizebox{8.5cm}{!}{\includegraphics[angle=270]{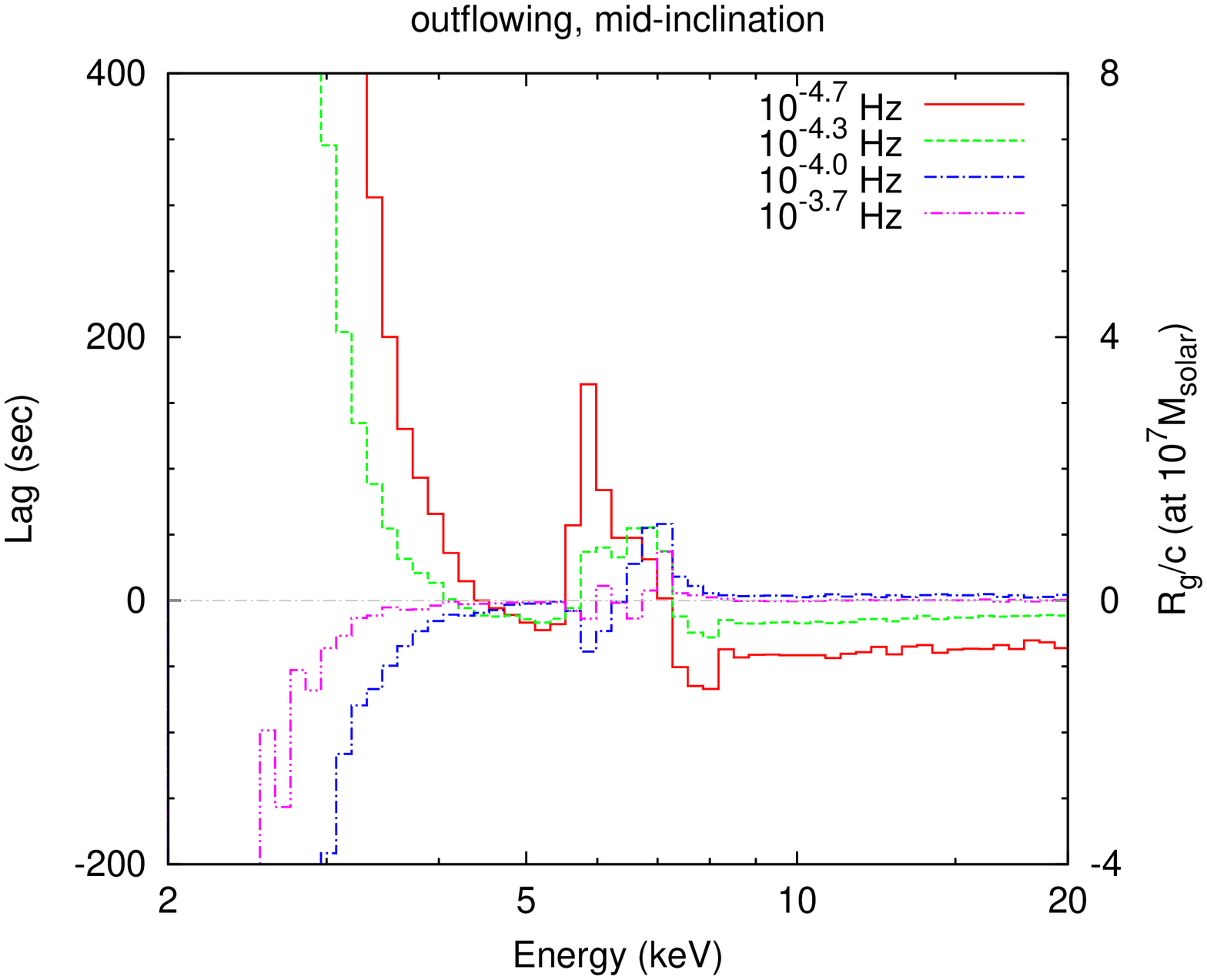}}
}
\subfigure{
\resizebox{8.5cm}{!}{\includegraphics[angle=270]{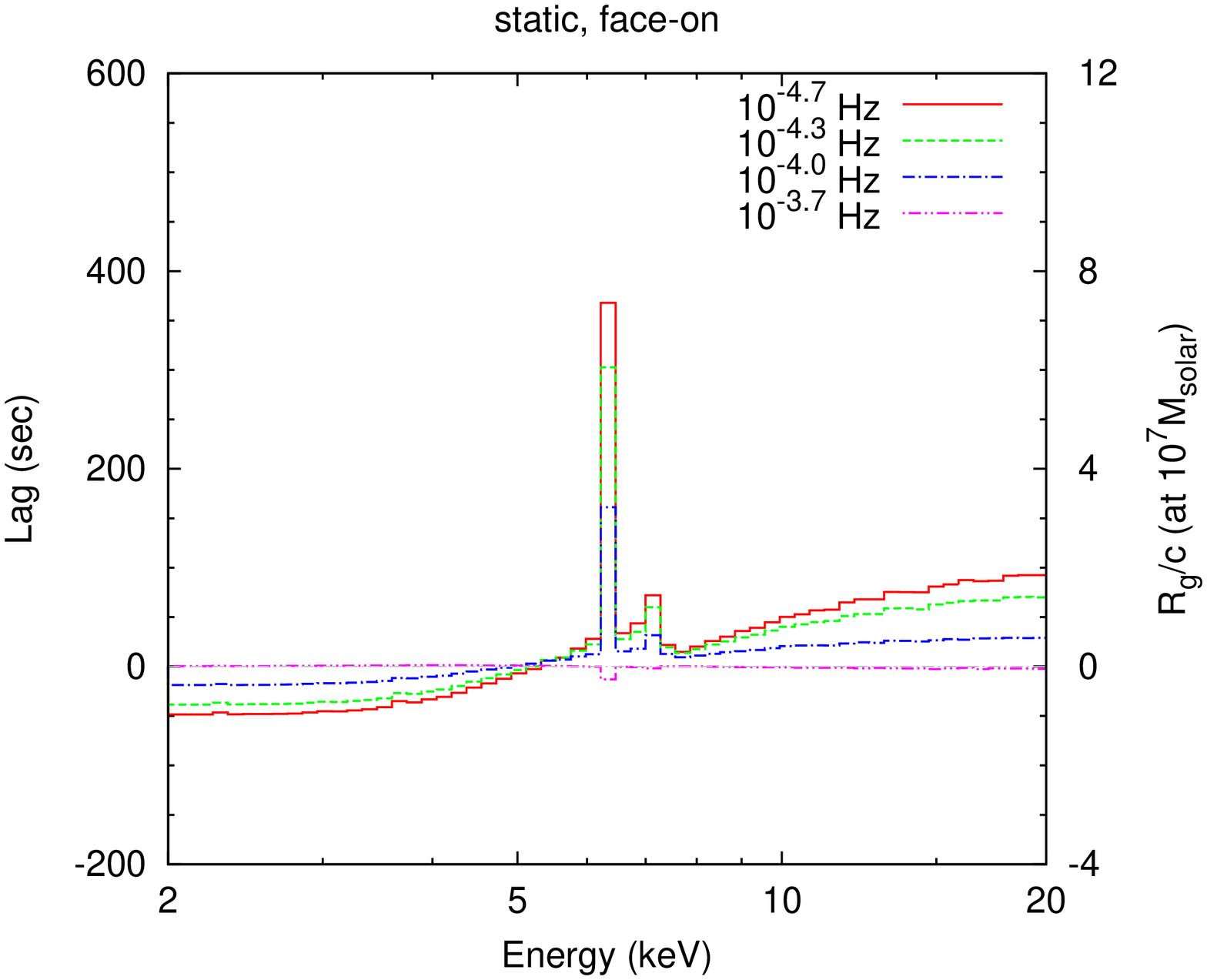}}
\resizebox{8.5cm}{!}{\includegraphics[angle=270]{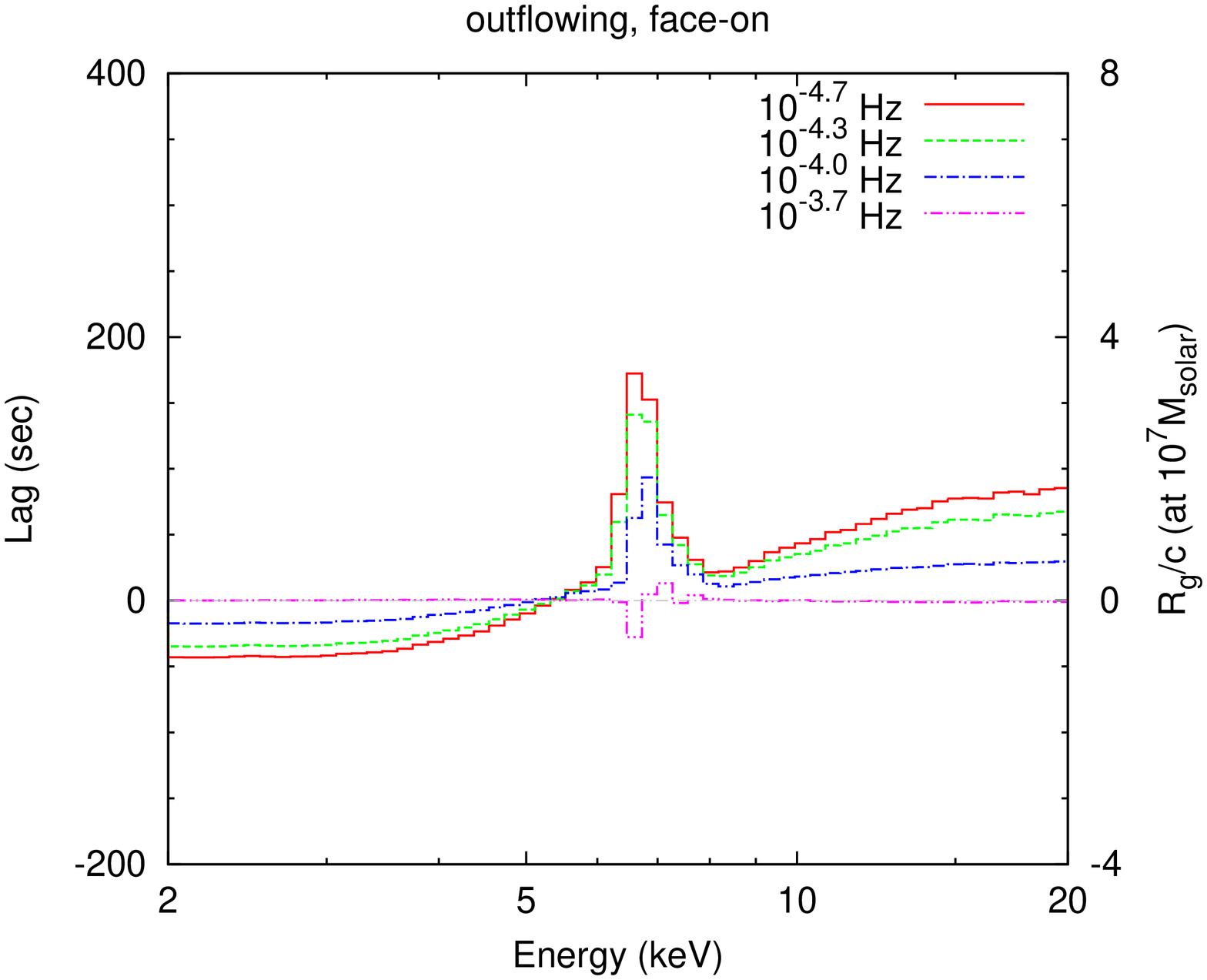}}
}
 \caption{Lag-energy plots at various frequencies derived from the Monte-Carlo simulation.
 }\label{fig:lagvsE}
 \end{center}
\end{figure*}

We use the same formalism to describe the lag-energy plots.  We use
each energy bin as a band of interest $h(t)$ and the entire 2--20~keV
(except for the energy bin of interest) as a reference band $s(t)$, to
get maximum statistics (e.g., \citealt{utt14,kar16}).  Fig.\
\ref{fig:lagvsE} shows the lag-energy plots derived over four
different frequencies to illustrate how this changes as a function of
variability timescale. 
The Fe-K lag timescale (as well as all other lag features) become
smaller as the frequency increases past $f_0=c/(2R)$ as the light
travel time smears out faster variability. For lower frequencies,
narrow Fe-K lines are seen from the static shell, while the line
feature is smaller in lag and broader in energy for the wind. 
The large lags are seen in the soft energy band because the primary photons are heavily absorbed by the neutral shell. Such large lags would disappear when we consider ionised materials.

In the low frequency limit ($f\to 0$) the lags as a function of energy
including the effect of dilution can be calculated
analytically. Appendix B1 shows that the lag as a function of energy,
$\tau(E)$, can written as 
\begin{equation}
\Delta \tau(E) = {\rm DF}(E) (f\rightarrow0)\Delta \tau
\approx \Bigl(\frac{R(E)}{P(E)} - \frac{R_{\rm tot}}{P_{\rm tot}}\Bigr)\Delta \tau
\label{eq:dilutionE}
\end{equation}
where $\Delta\tau$ is again the mean intrinsic lag of $R/c\sim
5000$~s. The measured lag times on the lowest frequency lag-energy
plots are indeed close to these values: at the iron line energy then
for the outflowing, edge on geometry this equation predicts that
$\Delta\tau(6.8~{\rm keV})\sim 0.029\times \Delta\tau\sim 150$~s (see
Appendix \ref{app2.1} and Fig.\ \ref{fig:DF_E}). 
The iron line feature is narrow in the static case, so the lags at the
iron line energy are
longer as the lagged line photons are concentrated in a single energy
bin so the effect of dilution is not so marked. 

At higher frequencies, the iron line lag becomes smaller as the fast
variations are smeared out by the light-travel time across the shell
(see Fig. \ref{fig:model}) in both the outflowing and static geometries, going
towards zero as the frequency increases towards $2\times 10^{-4}$~Hz.
For the static case, the lag-energy at frequencies approaching this
limit shows a simple decrease in amplitude around zero. However, for
the outflow, the line energy correlates with  $\phi$, which correlates
with the lag time. Hence the
centroid energy of the broad Fe-K line in the lag-energy plot shifts
depending on the variability timescale.  Long variability times are
required to see the longest lags, which come from material on the far
side of the shell, where the line is redshifted.  On the other hand,
material on the near side of the shell, where the line is blueshifted,
can respond to the shorter variability timescales.  Thus the
lag-energy plots for the outflowing shell have the line energy
shifting from red to blue as the variability timescale decreases,
correlated with decreasing line amplitude as the shell 
contributes less at higher frequencies. This is the first
demonstration that a wind can also produce changes in iron line
profile with Fourier frequency in lag-energy spectra.

\section{Discussion} \label{sec4}

Reverberation techniques were first used in AGN to try to constrain
the geometry of the broad line region. In this case the line can be
spectrally separated from the continuum, and the observed lags
directly show the difference of the light-travel time between the
illuminating source and the scattering materials (e.g.,
\citealt{pet93}).  However, in X-ray reverberation, both the continuum
and the lagged emission share the same energy band, and thus the
observed lag does not correspond to the light-travel time.  In fact,
Figs.\ \ref{fig:tf} and \ref{fig:lagvsf} show that the average delay
times of the simulated photons are much shorter than the light-travel
time to the scattering medium because the primary continuum photons
are dominant even in the Fe-K energy band.  In other words, the
primary photons ``dilute'' the lag amplitude of the reprocessed
photons, reducing the observed lags in both lag-frequency and
lag-energy plots. As was already stressed by several authors (e.g.,
\citealt{mil10a, kar13, utt14, gar14}), the observed lag time is not a
clean diagnostic of the size of the region.  Instead, it depends on
the dilution factor, which is given by the spectral decomposition, and
this is generally not unique. 
Dilution of the lag exists in both lag-frequency and lag-energy plots.
This dilution depends on the
amount of reprocessed emission in the reference band, as well as the
amount of primary emission in the lagged  band. Thus the short
lag times observed at the Fe-K line energy in several objects 
\citep{kar16} do not necessarily indicate
that this is produced very close to the event horizon of the black
hole. 

Instead, we have argued that the frequency at which
  the lags are attenuated is a more robust estimator of the size scale,
  with the reverberation lags being seen below $\sim c/R$~Hz.  Thus
  even the most extreme object, 1H 0707--495, where the Fe-K band lag
  is of order of 50~s, has reverberation lags which are detected up to
  $7\times 10^{-3}$~Hz (\citealt{kar13b}).  This indicates a firm
  upper limit to location of the reverberating region of $143$~s,
  i.e., $\sim 14\,R_g$.  This size scale is of the same order as the launching
  radius of a super-Eddington wind inferred from the data,
  $\sim20\,R_g$ \citep{don16}, and the observed broad profile in the
  lag-energy plot can be naturally explained by the Doppler shifts in
  the wind.

The equivalent width of Fe-K emission line in our calculation is $\sim50-100$~eV with Doppler broadening of $\sim0.7$~keV (for edge-on and mid-inclination) and $\sim0.3$~keV (for face-on) in the outflowing cases,
which is less than the observed Fe-K skewed emission line.
Whereas scattering on the shell cannot produce the observed Fe-K spectral feature,
the Fe-K absorption edge due to the shell may mimic the broad spectral feature if the shell is clumpy and partially covers the X-ray source, (e.g., \citealt{tan04,miz14}).
\citet{hag16} also argued that the blueshifted absorption line due to disc winds can mimic the spectral feature in 1H 0707--495.
In these manners, we suggest that the distant materials can explain both the energy spectra and the lag features in the Fe-K band simultaneously.
Here we do not consider disc reflection for simplicity, but of course,
disc within $100\,R_g$ must exist and affect the reverberation lags.
Reverberation of the Compton hump seen in some targets (e.g., \citealt{zog14,kar15a}) might trace such disc reflection,
because the hump expected by our picture is little.

The lag-energy plots show that the correlation of velocity with position as in
a radial outflow leads to shifts in the line energy and
amplitude. Such shifts are seen in NGC~4151 \citep{zog12}, where at
frequencies $\le 2\times 10^{-5}$~Hz there is a distinct feature
around the iron line energy, which becomes much broader, i.e., much
less distinct at higher frequencies (see Figs.\ 2 and 7 of
\citealt{zog12}). This is similar to the behaviour of the line in the
lag-energy spectra with an outflow shown in our Fig.\ \ref{fig:lagvsE}
(right column, middle panel).  The line response is strongly
suppressed on frequencies smaller than $c/R$, so the data from
NGC~4151 are then consistent with being from a wind at a few 100's of
$R_g$ rather than $\sim 5\,R_g$ as modelled by \citet{cac14}.

\section{Conclusion} \label{sec5}

We calculate the X-ray reverberation lags produced by a quite distant
matter around AGN via a Monte-Carlo simulation.  In our simulations,
the scattering material is assumed to be a neutral, part-spherical
thin shell at $100\,R_g$ with the black-hole mass of $10^7\,M_\odot$,
which can be either static or outflowing at $0.14c$.  We compiled
results for each simulation over three different inclination angles.

We found short-time ($\lesssim150$~s) hard lags 
between a continuum (3--4~keV) and Fe-K band (5--7~keV)  
in the frequency range of $\lesssim10^{-4}$~Hz, which is
much shorter than the light-travel time (5000~s).  
The short lag amplitude compared to
the light-travel time is explained by the dilution effect, where
the majority of the photons in the Fe-K energy band are primary
photons without time-delay, whereas the time-delayed reprocessed photons
make a subtle contribution. 
The line is broadened in the lag-energy plots when the scattering
material is an outflowing wind;
photons scattered on the near side are blueshifted, 
whereas those on the far side are redshifted. This makes a 
broad feature with short lag time in lag-energy plots, 
similar to that seen in 1H0707-495. The correlation of 
line of sight velocity in the wind and lag time leads to shifts in the
observed energy of the line as a function of 
frequency, similar to those seen in NGC~4151 \citep{zog12}.
Hence we show that the observed X-ray reverberation lag features can
be produced by the outflowing matter such as disc winds at
$\sim 100\,R_g$.
This is a viable alternative geometry to a compact corona
close to the event horizon of a high spin black hole.

\section*{Acknowledgements}
Authors are financially supported by the JSPS/MEXT KAKENHI Grant Numbers JP15J07567 (MM), JP16K05309 (KE), and JP24105007, JP15H03642, JP16K05309 (MT).
CD acknowledges STFC funding under grant ST/L00075X/1 and a JSPS long term fellowship L16581.







\bibliographystyle{mnras}
\bibliography{mn-jour,00}

\begin{thebibliography}{}
\makeatletter
\relax
\def\mn@urlcharsother{\let\do\@makeother \do\$\do\&\do\#\do\^\do\_\do\%\do\~}
\def\mn@doi{\begingroup\mn@urlcharsother \@ifnextchar [ {\mn@doi@}
  {\mn@doi@[]}}
\def\mn@doi@[#1]#2{\def\@tempa{#1}\ifx\@tempa\@empty \href
  {http://dx.doi.org/#2} {doi:#2}\else \href {http://dx.doi.org/#2} {#1}\fi
  \endgroup}
\def\mn@eprint#1#2{\mn@eprint@#1:#2::\@nil}
\def\mn@eprint@arXiv#1{\href {http://arxiv.org/abs/#1} {{\tt arXiv:#1}}}
\def\mn@eprint@dblp#1{\href {http://dblp.uni-trier.de/rec/bibtex/#1.xml}
  {dblp:#1}}
\def\mn@eprint@#1:#2:#3:#4\@nil{\def\@tempa {#1}\def\@tempb {#2}\def\@tempc
  {#3}\ifx \@tempc \@empty \let \@tempc \@tempb \let \@tempb \@tempa \fi \ifx
  \@tempb \@empty \def\@tempb {arXiv}\fi \@ifundefined
  {mn@eprint@\@tempb}{\@tempb:\@tempc}{\expandafter \expandafter \csname
  mn@eprint@\@tempb\endcsname \expandafter{\@tempc}}}

\bibitem[\protect\citeauthoryear{{Agostinelli} et~al.,}{{Agostinelli}
  et~al.}{2003}]{ago03}
{Agostinelli} S.,  et~al., 2003, \mn@doi [Nuclear Instruments and Methods in
  Physics Research A] {10.1016/S0168-9002(03)01368-8}, \href
  {http://adsabs.harvard.edu/abs/2003NIMPA.506..250A} {506, 250}

\bibitem[\protect\citeauthoryear{{Allison} et~al.,}{{Allison}
  et~al.}{2006}]{all06}
{Allison} J.,  et~al., 2006, \mn@doi [IEEE Transactions on Nuclear Science]
  {10.1109/TNS.2006.869826}, \href
  {http://adsabs.harvard.edu/abs/2006ITNS...53..270A} {53, 270}

\bibitem[\protect\citeauthoryear{{Ar{\'e}valo} \& {Uttley}}{{Ar{\'e}valo} \&
  {Uttley}}{2006}]{are06}
{Ar{\'e}valo} P.,  {Uttley} P.,  2006, \mn@doi [\mnras]
  {10.1111/j.1365-2966.2006.09989.x}, \href
  {http://adsabs.harvard.edu/abs/2006MNRAS.367..801A} {367, 801}

\bibitem[\protect\citeauthoryear{{Cackett}, {Zoghbi}, {Reynolds}, {Fabian},
  {Kara}, {Uttley}  \& {Wilkins}}{{Cackett} et~al.}{2014}]{cac14}
{Cackett} E.~M.,  {Zoghbi} A.,  {Reynolds} C.,  {Fabian} A.~C.,  {Kara} E.,
  {Uttley} P.,   {Wilkins} D.~R.,  2014, \mn@doi [\mnras]
  {10.1093/mnras/stt2424}, \href
  {http://adsabs.harvard.edu/abs/2014MNRAS.438.2980C} {438, 2980}

\bibitem[\protect\citeauthoryear{{Chainakun}, {Young}  \& {Kara}}{{Chainakun}
  et~al.}{2016}]{cha16}
{Chainakun} P.,  {Young} A.~J.,   {Kara} E.,  2016, \mn@doi [\mnras]
  {10.1093/mnras/stw1105}, \href
  {http://adsabs.harvard.edu/abs/2016MNRAS.460.3076C} {460, 3076}

\bibitem[\protect\citeauthoryear{{Done} \& {Jin}}{{Done} \&
  {Jin}}{2016}]{don16}
{Done} C.,  {Jin} C.,  2016, \mn@doi [\mnras] {10.1093/mnras/stw1070}, \href
  {http://adsabs.harvard.edu/abs/2016MNRAS.460.1716D} {460, 1716}

\bibitem[\protect\citeauthoryear{{Epitropakis}, {Papadakis}, {Dov{\v c}iak},
  {Pech{\'a}{\v c}ek}, {Emmanoulopoulos}, {Karas}  \& {McHardy}}{{Epitropakis}
  et~al.}{2016}]{epi16}
{Epitropakis} A.,  {Papadakis} I.~E.,  {Dov{\v c}iak} M.,  {Pech{\'a}{\v c}ek}
  T.,  {Emmanoulopoulos} D.,  {Karas} V.,   {McHardy} I.~M.,  2016, \mn@doi
  [\aap] {10.1051/0004-6361/201527748}, \href
  {http://adsabs.harvard.edu/abs/2016A%26A...594A..71E} {594, A71}

\bibitem[\protect\citeauthoryear{{Fabian} et~al.,}{{Fabian}
  et~al.}{2009}]{fab09}
{Fabian} A.~C.,  et~al., 2009, \mn@doi [\nat] {10.1038/nature08007}, \href
  {http://adsabs.harvard.edu/abs/2009Natur.459..540F} {459, 540}

\bibitem[\protect\citeauthoryear{{Gardner} \& {Done}}{{Gardner} \&
  {Done}}{2014}]{gar14}
{Gardner} E.,  {Done} C.,  2014, \mn@doi [\mnras] {10.1093/mnras/stu1026},
  \href {http://adsabs.harvard.edu/abs/2014MNRAS.442.2456G} {442, 2456}

\bibitem[\protect\citeauthoryear{{Gardner} \& {Done}}{{Gardner} \&
  {Done}}{2015}]{gar15}
{Gardner} E.,  {Done} C.,  2015, \mn@doi [\mnras] {10.1093/mnras/stv168}, \href
  {http://adsabs.harvard.edu/abs/2015MNRAS.448.2245G} {448, 2245}

\bibitem[\protect\citeauthoryear{{Hagino}, {Odaka}, {Done}, {Gandhi},
  {Watanabe}, {Sako}  \& {Takahashi}}{{Hagino} et~al.}{2015}]{hag15}
{Hagino} K.,  {Odaka} H.,  {Done} C.,  {Gandhi} P.,  {Watanabe} S.,  {Sako} M.,
    {Takahashi} T.,  2015, \mn@doi [\mnras] {10.1093/mnras/stu2095}, \href
  {http://adsabs.harvard.edu/abs/2015MNRAS.446..663H} {446, 663}

\bibitem[\protect\citeauthoryear{{Hagino}, {Odaka}, {Done}, {Tomaru},
  {Watanabe}  \& {Takahashi}}{{Hagino} et~al.}{2016}]{hag16}
{Hagino} K.,  {Odaka} H.,  {Done} C.,  {Tomaru} R.,  {Watanabe} S.,
  {Takahashi} T.,  2016, \mn@doi [\mnras] {10.1093/mnras/stw1579}, \href
  {http://adsabs.harvard.edu/abs/2016MNRAS.461.3954H} {461, 3954}

\bibitem[\protect\citeauthoryear{{Kara}, {Fabian}, {Cackett}, {Steiner},
  {Uttley}, {Wilkins}  \& {Zoghbi}}{{Kara} et~al.}{2013a}]{kar13b}
{Kara} E.,  {Fabian} A.~C.,  {Cackett} E.~M.,  {Steiner} J.~F.,  {Uttley} P.,
  {Wilkins} D.~R.,   {Zoghbi} A.,  2013a, \mn@doi [\mnras]
  {10.1093/mnras/sts155}, \href
  {http://adsabs.harvard.edu/abs/2013MNRAS.428.2795K} {428, 2795}

\bibitem[\protect\citeauthoryear{{Kara}, {Fabian}, {Cackett}, {Uttley},
  {Wilkins}  \& {Zoghbi}}{{Kara} et~al.}{2013b}]{kar13}
{Kara} E.,  {Fabian} A.~C.,  {Cackett} E.~M.,  {Uttley} P.,  {Wilkins} D.~R.,
  {Zoghbi} A.,  2013b, \mn@doi [\mnras] {10.1093/mnras/stt1055}, \href
  {http://adsabs.harvard.edu/abs/2013MNRAS.434.1129K} {434, 1129}

\bibitem[\protect\citeauthoryear{{Kara}, {Cackett}, {Fabian}, {Reynolds}  \&
  {Uttley}}{{Kara} et~al.}{2014}]{kar14}
{Kara} E.,  {Cackett} E.~M.,  {Fabian} A.~C.,  {Reynolds} C.,   {Uttley} P.,
  2014, \mn@doi [\mnras] {10.1093/mnrasl/slt173}, \href
  {http://adsabs.harvard.edu/abs/2014MNRAS.439L..26K} {439, L26}

\bibitem[\protect\citeauthoryear{{Kara} et~al.,}{{Kara} et~al.}{2015}]{kar15a}
{Kara} E.,  et~al., 2015, \mn@doi [\mnras] {10.1093/mnras/stu2136}, \href
  {http://adsabs.harvard.edu/abs/2015MNRAS.446..737K} {446, 737}

\bibitem[\protect\citeauthoryear{{Kara}, {Alston}, {Fabian}, {Cackett},
  {Uttley}, {Reynolds}  \& {Zoghbi}}{{Kara} et~al.}{2016}]{kar16}
{Kara} E.,  {Alston} W.~N.,  {Fabian} A.~C.,  {Cackett} E.~M.,  {Uttley} P.,
  {Reynolds} C.~S.,   {Zoghbi} A.,  2016, \mn@doi [\mnras]
  {10.1093/mnras/stw1695}, \href
  {http://adsabs.harvard.edu/abs/2016MNRAS.462..511K} {462, 511}

\bibitem[\protect\citeauthoryear{{Kotov}, {Churazov}  \& {Gilfanov}}{{Kotov}
  et~al.}{2001}]{kot01}
{Kotov} O.,  {Churazov} E.,   {Gilfanov} M.,  2001, \mn@doi [\mnras]
  {10.1046/j.1365-8711.2001.04769.x}, \href
  {http://adsabs.harvard.edu/abs/2001MNRAS.327..799K} {327, 799}

\bibitem[\protect\citeauthoryear{{Miller}, {Turner}, {Reeves}, {Lobban},
  {Kraemer}  \& {Crenshaw}}{{Miller} et~al.}{2010a}]{mil10a}
{Miller} L.,  {Turner} T.~J.,  {Reeves} J.~N.,  {Lobban} A.,  {Kraemer} S.~B.,
   {Crenshaw} D.~M.,  2010a, \mn@doi [\mnras]
  {10.1111/j.1365-2966.2009.16149.x}, \href
  {http://adsabs.harvard.edu/abs/2010MNRAS.403..196M} {403, 196}

\bibitem[\protect\citeauthoryear{{Miller}, {Turner}, {Reeves}  \&
  {Braito}}{{Miller} et~al.}{2010b}]{mil10b}
{Miller} L.,  {Turner} T.~J.,  {Reeves} J.~N.,   {Braito} V.,  2010b, \mn@doi
  [\mnras] {10.1111/j.1365-2966.2010.17261.x}, \href
  {http://adsabs.harvard.edu/abs/2010MNRAS.408.1928M} {408, 1928}

\bibitem[\protect\citeauthoryear{{Mizumoto}, {Ebisawa}  \&
  {Sameshima}}{{Mizumoto} et~al.}{2014}]{miz14}
{Mizumoto} M.,  {Ebisawa} K.,   {Sameshima} H.,  2014, \mn@doi [\pasj]
  {10.1093/pasj/psu121}, \href
  {http://adsabs.harvard.edu/abs/2014PASJ...66..122M} {66, 122}

\bibitem[\protect\citeauthoryear{{Mizumoto}, {Moriyama}, {Ebisawa},
  {Mineshige}, {Kawanaka}  \& {Tsujimoto}}{{Mizumoto} et~al.}{2018}]{miz18a}
{Mizumoto} M.,  {Moriyama} K.,  {Ebisawa} K.,  {Mineshige} S.,  {Kawanaka} N.,
   {Tsujimoto} M.,  2018, \pasj, \href
  {http://adsabs.harvard.edu/abs/2018arXiv180207554M} {accepted (arXiv:
  1802.07554)}

\bibitem[\protect\citeauthoryear{{Nied{\'z}wiecki}, {Zdziarski}  \&
  {Szanecki}}{{Nied{\'z}wiecki} et~al.}{2016}]{nie16}
{Nied{\'z}wiecki} A.,  {Zdziarski} A.~A.,   {Szanecki} M.,  2016, \mn@doi
  [\apjl] {10.3847/2041-8205/821/1/L1}, \href
  {http://adsabs.harvard.edu/abs/2016ApJ...821L...1N} {821, L1}

\bibitem[\protect\citeauthoryear{{Nowak}, {Vaughan}, {Wilms}, {Dove}  \&
  {Begelman}}{{Nowak} et~al.}{1999}]{now99}
{Nowak} M.~A.,  {Vaughan} B.~A.,  {Wilms} J.,  {Dove} J.~B.,   {Begelman}
  M.~C.,  1999, \mn@doi [\apj] {10.1086/306610}, \href
  {http://adsabs.harvard.edu/abs/1999ApJ...510..874N} {510, 874}

\bibitem[\protect\citeauthoryear{{Odaka}, {Aharonian}, {Watanabe}, {Tanaka},
  {Khangulyan}  \& {Takahashi}}{{Odaka} et~al.}{2011}]{oda11}
{Odaka} H.,  {Aharonian} F.,  {Watanabe} S.,  {Tanaka} Y.,  {Khangulyan} D.,
  {Takahashi} T.,  2011, \mn@doi [\apj] {10.1088/0004-637X/740/2/103}, \href
  {http://adsabs.harvard.edu/abs/2011ApJ...740..103O} {740, 103}

\bibitem[\protect\citeauthoryear{{Peterson}}{{Peterson}}{1993}]{pet93}
{Peterson} B.~M.,  1993, \mn@doi [\pasp] {10.1086/133140}, \href
  {http://adsabs.harvard.edu/abs/1993PASP..105..247P} {105, 247}

\bibitem[\protect\citeauthoryear{{Tanaka}, {Boller}, {Gallo}, {Keil}  \&
  {Ueda}}{{Tanaka} et~al.}{2004}]{tan04}
{Tanaka} Y.,  {Boller} T.,  {Gallo} L.,  {Keil} R.,   {Ueda} Y.,  2004, \mn@doi
  [\pasj] {10.1093/pasj/56.3.L9}, \href
  {http://adsabs.harvard.edu/abs/2004PASJ...56L...9T} {56, L9}

\bibitem[\protect\citeauthoryear{{Tombesi}, {Cappi}, {Reeves}, {Palumbo},
  {Braito}  \& {Dadina}}{{Tombesi} et~al.}{2011}]{tom11}
{Tombesi} F.,  {Cappi} M.,  {Reeves} J.~N.,  {Palumbo} G.~G.~C.,  {Braito} V.,
   {Dadina} M.,  2011, \mn@doi [\apj] {10.1088/0004-637X/742/1/44}, \href
  {http://adsabs.harvard.edu/abs/2011ApJ...742...44T} {742, 44}

\bibitem[\protect\citeauthoryear{{Turner}, {Miller}, {Reeves}  \&
  {Braito}}{{Turner} et~al.}{2017}]{tur17}
{Turner} T.~J.,  {Miller} L.,  {Reeves} J.~N.,   {Braito} V.,  2017, \mn@doi
  [\mnras] {10.1093/mnras/stx388}, \href
  {http://adsabs.harvard.edu/abs/2017MNRAS.467.3924T} {467, 3924}

\bibitem[\protect\citeauthoryear{{Uttley}, {Cackett}, {Fabian}, {Kara}  \&
  {Wilkins}}{{Uttley} et~al.}{2014}]{utt14}
{Uttley} P.,  {Cackett} E.~M.,  {Fabian} A.~C.,  {Kara} E.,   {Wilkins} D.~R.,
  2014, \mn@doi [\aapr] {10.1007/s00159-014-0072-0}, \href
  {http://adsabs.harvard.edu/abs/2014A%26ARv..22...72U} {22, 72}

\bibitem[\protect\citeauthoryear{{Vaughan} \& {Nowak}}{{Vaughan} \&
  {Nowak}}{1997}]{vau97}
{Vaughan} B.~A.,  {Nowak} M.~A.,  1997, \mn@doi [\apjl] {10.1086/310430}, \href
  {http://adsabs.harvard.edu/abs/1997ApJ...474L..43V} {474, L43}

\bibitem[\protect\citeauthoryear{{Wilkins}, {Cackett}, {Fabian}  \&
  {Reynolds}}{{Wilkins} et~al.}{2016}]{wil16}
{Wilkins} D.~R.,  {Cackett} E.~M.,  {Fabian} A.~C.,   {Reynolds} C.~S.,  2016,
  \mn@doi [\mnras] {10.1093/mnras/stw276}, \href
  {http://adsabs.harvard.edu/abs/2016MNRAS.458..200W} {458, 200}

\bibitem[\protect\citeauthoryear{{Zhou} \& {Wang}}{{Zhou} \&
  {Wang}}{2005}]{zho05}
{Zhou} X.-L.,  {Wang} J.-M.,  2005, \mn@doi [\apjl] {10.1086/427871}, \href
  {http://adsabs.harvard.edu/abs/2005ApJ...618L..83Z} {618, L83}

\bibitem[\protect\citeauthoryear{{Zoghbi}, {Uttley}  \& {Fabian}}{{Zoghbi}
  et~al.}{2011}]{zog11}
{Zoghbi} A.,  {Uttley} P.,   {Fabian} A.~C.,  2011, \mn@doi [\mnras]
  {10.1111/j.1365-2966.2010.17883.x}, \href
  {http://adsabs.harvard.edu/abs/2011MNRAS.412...59Z} {412, 59}

\bibitem[\protect\citeauthoryear{{Zoghbi}, {Fabian}, {Reynolds}  \&
  {Cackett}}{{Zoghbi} et~al.}{2012}]{zog12}
{Zoghbi} A.,  {Fabian} A.~C.,  {Reynolds} C.~S.,   {Cackett} E.~M.,  2012,
  \mn@doi [\mnras] {10.1111/j.1365-2966.2012.20587.x}, \href
  {http://adsabs.harvard.edu/abs/2012MNRAS.422..129Z} {422, 129}

\bibitem[\protect\citeauthoryear{{Zoghbi} et~al.,}{{Zoghbi}
  et~al.}{2014}]{zog14}
{Zoghbi} A.,  et~al., 2014, \mn@doi [\apj] {10.1088/0004-637X/789/1/56}, \href
  {http://adsabs.harvard.edu/abs/2014ApJ...789...56Z} {789, 56}

\makeatother
\end{thebibliography}


\appendix

\section{Response of a top-hat function} \label{app1}

In \S\ref{sec:lagf}, we derived equation (\ref{eq:dilution}) assuming that the reprocessed component can be represented by 
a single intrinsic delay time, or the response is a delta-function.
However, this assumption is too simplified, because the response function (see bottom panels in Fig.\ \ref{fig:tf}) is not like a delta-function; it is rather like a top-hat function.
In this section, we analytically evaluate the lags with response of the top-hat function.

\begin{figure}
  \begin{center}
\includegraphics[angle=270,width=\columnwidth]{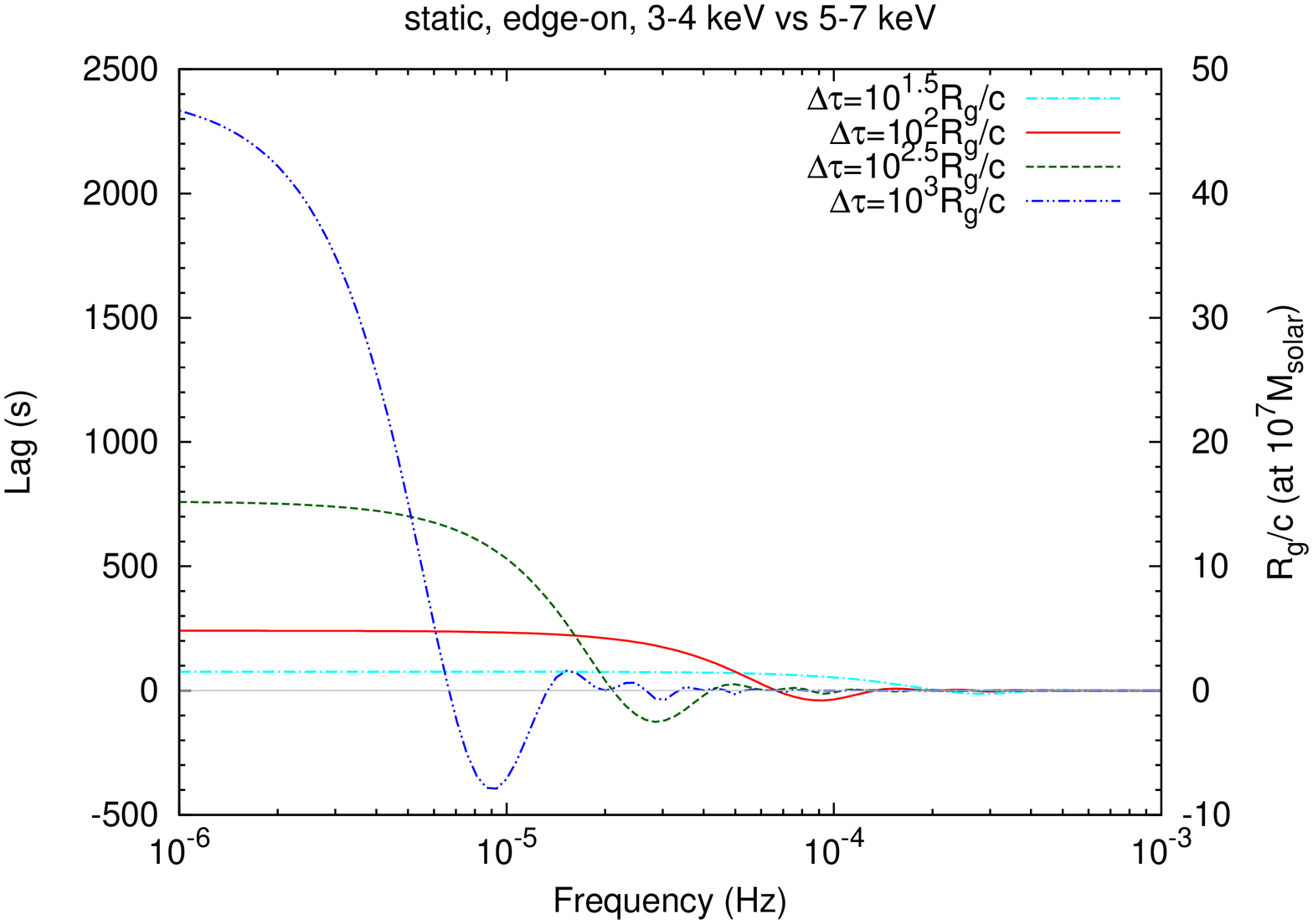}
  \caption{
Same as Fig.\ \ref{fig:model}, but the top-hat function is assumed ($\tau_{\rm min}=0$ and $\tau_{\rm max}=2\Delta \tau$ in equation \ref{eq:tophat}). 
}\label{fig:model2}
  \end{center}
\end{figure}

First, we rewrite equation (\ref{eq:ltcrv}) as
\begin{equation}
\begin{split}
s(t)&=P_sg(t)+R_s\int\!\! dt^\prime \psi_s(t^\prime)g(t-t^\prime)\\
h(t)&=P_hg(t)+R_h\int\!\! dt^\prime \psi_h(t^\prime)g(t-t^\prime),
\end{split}
\end{equation}
where $R_s$ and $R_h$ show the reprocessed components in the soft- and hard-band, and
$\psi_s$ and $\psi_h$ are normalised response functions ($\int\!\!dt\,\psi_s(t)=\int\!\!dt\,\psi_h(t)=1$).
In this case, 
\begin{equation}
\mathcal{S}(f)\mathcal{H}^*(f)
= \{P_s+R_s\Psi_s(f)\}\{P_h+R_h\Psi^*_h(f)\} |\mathcal{G}(f)|^2,
\end{equation}
where $\Psi_s(f)$ and $\Psi_h(f)$ is a Fourier transform of $\psi_s(t)$ and $\psi_h(t)$.
Under the assumption that $\psi_s(t)=\psi_h(t)=\psi(t)$, or $\Psi_s(f)=\Psi_h(f)=\Psi(f)=a(f)+ib(f)$,
$\tau(f)$ is calculated as
\begin{equation}
\tau(f)= \frac{1}{2\pi f}\arctan\left(\frac{-b(P_sR_h-R_sP_h)}{P_sP_h+R_sR_h+a(P_sR_h+R_sP_h)}\right). 
\end{equation}
When the response is a top-hat function,
\begin{equation}
\psi(t)=\begin{cases}
1/\Delta \tau & (\tau_{\rm min} \leq t \leq \tau_{\rm max})\\
0 & (\mathrm{otherwise}),
\end{cases}
\end{equation}
$\Psi(f)$ is written as
\begin{align}
\Psi(f)&=\frac{1}{-2\pi if\Delta \tau}\left( \exp[-2\pi if\tau_{\rm max}]-\exp[-2\pi if\tau_{\rm min}] \right) \nonumber \\
&=\frac{1}{2\pi f\Delta \tau}\left[\left\{2\cos \left(2\pi f \tau_{\rm ave}\right)\sin \left(2\pi f \frac{\Delta \tau}{2} \right) \right\} \right. \nonumber\\
&  \:\:\:\:\:\:\:\:\:\:\:\:\:\:\:\:\:\:\:\:\:\:
\left.-i\left\{2\sin \left(2\pi f \tau_{\rm ave}\right)\sin\left( 2\pi f \frac{\Delta \tau}{2} \right)\right\}\right], \label{eq:tophat}
\end{align}
where $\tau_{\rm ave}=(\tau_{\rm max}+\tau_{\rm min})/2$ and $\Delta \tau=(\tau_{\rm max}-\tau_{\rm min})/2$. 
In a limit of $f\rightarrow0$, we see lags on the longest timescale, so a finite width of the top-hat function can be ignored and
the response can be treated as a delta function, and
we can use the same dilution factor (equation \ref{eq:dilution2}).
Fig.\ \ref{fig:model2} illustrates the light-travel time dependence of the lag frequencies,
with the top-hat function ($\tau_{\rm min}=0$ and $\tau_{\rm max}=2\Delta \tau$).
The lags oscillate around zero at high frequencies, and almost disappear on $c/R$, which is same as the delta-function case (Fig.\ \ref{fig:model}).

\section{Details of the dilution effect}\label{app2}
\subsection{Dilution factor as a function of energy}\label{app2.1}

\begin{figure}
  \begin{center}
\includegraphics[angle=270,width=\columnwidth]{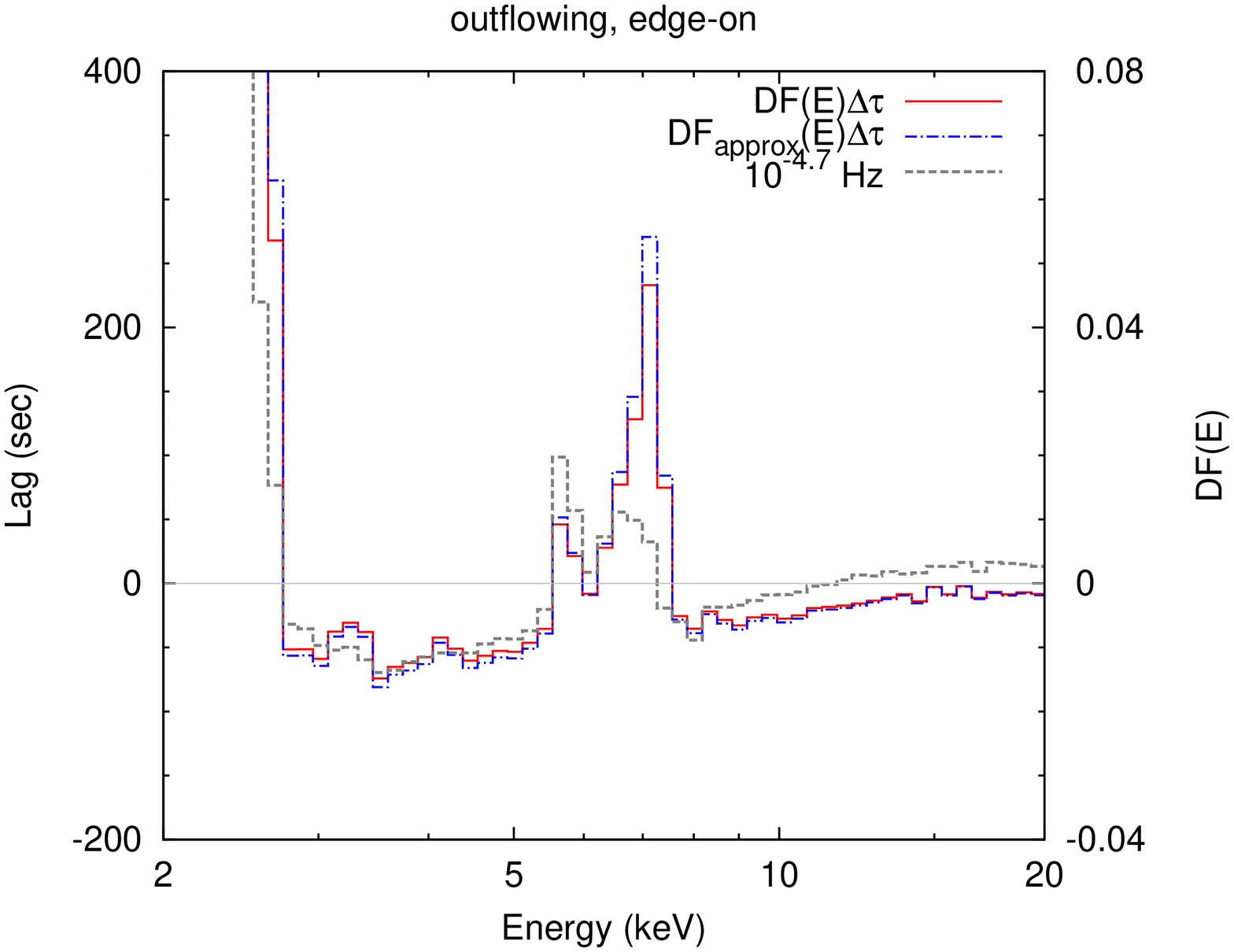}
  \caption{
The plot of ${\rm DF}(E)\Delta\tau$ (red-solid, shown by equation \ref{eq:DFE}), 
${\rm DF_{approx}(E)\Delta\tau}$ (blue-dot-dashed, shown by equation \ref{eq:DFE2}), and
the lag-energy plot in the outflowing edge-on case at $10^{-4.7}$~Hz (grey-dashed), which is same as the one in the right upper panel of Fig.\ \ref{fig:lagvsE}).
The right axis shows the ${\rm DF}(E)$ values.
}\label{fig:DF_E}
  \end{center}
\end{figure}

The dilution factor is derived in equation (\ref{eq:dilution2}) in the $f\to0$ limit.
The energy-dependent dilution factor ${\rm DF}(E)$ is similarly expressed as
\begin{equation}
\mathrm{DF}(E)
=\frac{P_{\rm tot}R(E)-R_{\rm tot}P(E)}{(P_{\rm tot}+R_{\rm tot})(P(E)+R(E))}, \label{eq:DFE}
\end{equation}
where 
$P(E)$ and $R(E)$ are the primary and reprocessed components in the energy bin of interest, and
$P_{\rm tot}$ and $R_{\rm tot}$ are those in the total energy band, like equation (\ref{eq:dilution}).
Under the assumption that $R_{\rm tot}/P_{\rm tot}\ll 1$ and  $R(E)/P(E)\ll 1$, equation (\ref{eq:DFE}) is approximated as
\begin{equation}
\mathrm{DF}_{\rm approx}(E)=\frac{R(E)}{P(E)}-\frac{R_{\rm tot}}{P_{\rm tot}} \label{eq:DFE2}.
\end{equation}
In other words, ${\rm DF}_{\rm approx}$ is calculated as difference between relative flux of the reprocessed component to the primary component in the energy bin of interest and that in the total energy band.
These equations work very well to describe the lag-energy spectrum 
seen from the static shell at low frequencies. 
Fig.\ \ref{fig:DF_E} shows comparison of the lag-energy plot 
in the low-frequency range with $\mathrm{DF}(E)$.  
We can see that 
$\tau_{f\to 0}(E)\simeq{\rm DF}(E)\Delta\tau\simeq {\rm DF}_{\rm approx}(E) \Delta \tau$, 
with only several differences around the line profile.

In all the cases, both static and outflowing, the Fe-K lag amplitudes
correspond to several $R_g/c$, much shorter than $R/c=100\,R_g/c$.
Calculating this explicitly for the outflowing case, 
at $E=6.8\,{\rm keV}$ (slightly lower energy 
than the blue peak of the outflowing line) gives
\begin{align}
\mathrm{DF}_{\rm approx}(E=6.8\,{\rm keV})&=
\frac{R(E)}{P(E)}-\frac{R_{\rm tot}}{P_{\rm tot}} \nonumber\\
&= 0.081-0.052 \nonumber\\
&= 0.029,
\end{align}
which means that the lag amplitude is diluted by almost two orders of
magnitude (see Fig~\ref{fig:DF_E}). The lag at the iron line energy in
the lag-energy spectrum can be as affected by dilution as is the
broader energy band used for the lag-frequency spectrum. 

\subsection{Separating lagged and primary emission}\label{app2.2}

\begin{figure}
  \begin{center}
\includegraphics[angle=270,width=\columnwidth]{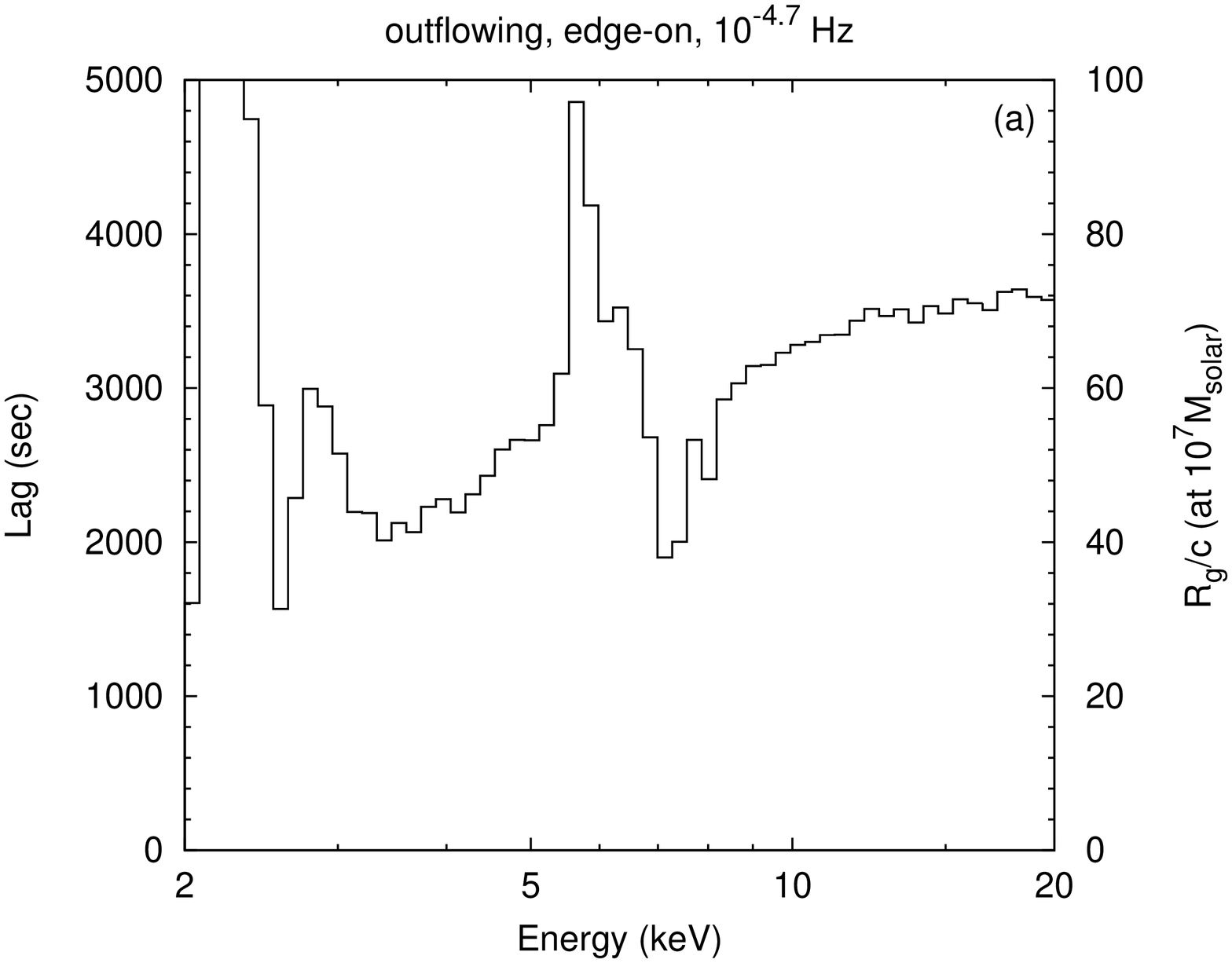}\\
\includegraphics[angle=270,width=\columnwidth]{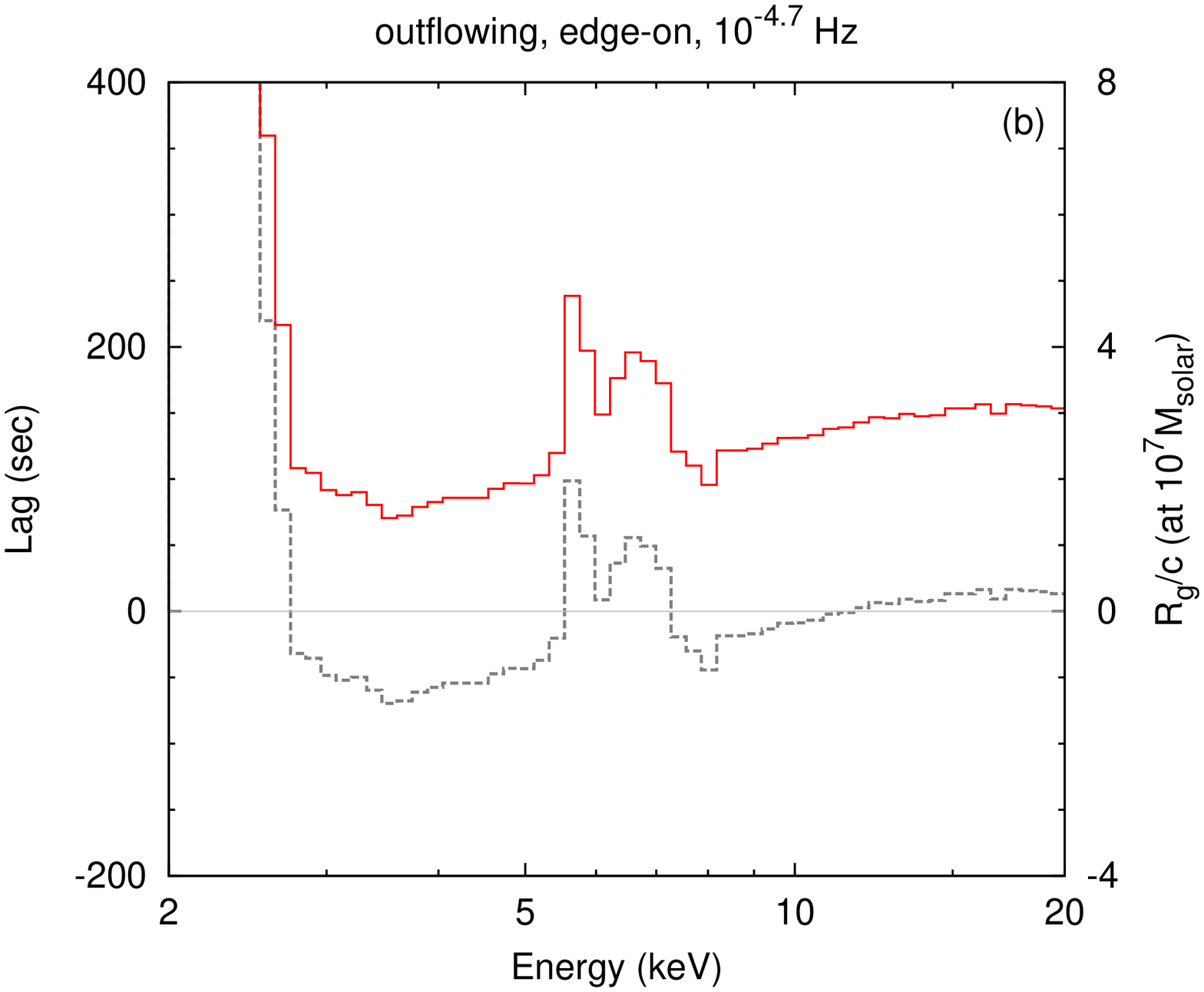}
  \caption{ 
(a) The lag-energy plot when the reference band has only the primary component and the energy bin of interest has only the reprocessed component in the outflowing edge-on case at $10^{-4.7}$~Hz, derived from the Monte-Carlo simulation.
(b) The one when the reference band has only the primary component but the energy bin of interest has both the primary and reprocessed components (red-solid).
The grey dashed line is same as the one in Fig.\ \ref{fig:DF_E}.
}\label{fig:type}
  \end{center}
\end{figure}

We have studied the dilution effect in the low-frequency limit in the above subsection (e.g., equation \ref{eq:DFE2}).
In this subsection, we investigate the dilution effect in all the frequency ranges.
We explained that lags are diluted because the primary and reprocessed components share the same energy band, so
here we calculate lags when they do not share the same band.
Since this is a simulation, we can separate out the lagged and primary emission. 
In an ideal situation, first, let us consider that
a reference band has only primary emission, whereas an energy band of interest has the only reprocessed emission.
In this case, the lag amplitudes are written as 
\begin{equation}
\tau_f(E)=\frac{1}{2\pi f}{\rm arg}\left[ P_{\rm tot}\sum_kR(E,k)\exp[2\pi ikt_{\rm bin}f] \right]. \nonumber\\
\end{equation}
The resultant lag-energy plot is shown in Fig.~\ref{fig:type}(a).
This plot is not affected by dilution and the lag amplitude is an order of $R/c=5000$~s, as expected,
but it does have energy dependence due to the lag weighting of the numerator.  
We investigate this in the low-frequency limit, where
\begin{align}
\tau_f(E)
&\xrightarrow[f\to 0]{}\frac{1}{2\pi f} {\rm arg}\left[ \sum_k R(E,k) +iR(E,k) 2\pi kt_{\rm bin}f \right] \nonumber\\
&=\frac{1}{2\pi f} \arctan\left(\frac{\sum_kR(E,k)2\pi kt_{\rm bin}f}{\sum_kR(E,k)}\right) \nonumber\\
&\to\frac{\sum_k kt_{\rm bin}R(E,k)}{\sum_kR(E,k)}. \label{eq:type1}
\end{align}
In effect, the difference between the reprocessed emission lagged by $kt_{\rm bin}$ and
the time averaged lagged spectrum produces the energy dependence.

Next, we explore the situation where the energy bin of interest contains the primary component as well as the reprocessed
one.
This has a very similar profile to the measured lag-energy plot (Fig.\ \ref{fig:type}b),
but the dilution effect takes place.
In the low-frequency limit, this plot is expressed as
\begin{equation}
\tau_{f}(E)\xrightarrow[f\to 0]{}\frac{\sum_k kt_{\rm bin}R(E,k)}{P(E)+\sum_kR(E,k)}. \label{eq:type2}
\end{equation}
We can see that the $P(E)$ component in the denominator acts as a diluting factor.  
The plot is shifted by $\sim150$~s because we ignored the reprocessed component in the reference band.  
Indeed, the effect of $R_{\rm tot}$ is estimated as $(R_{\rm tot}/P_{\rm tot})\tau$ from equation (\ref{eq:DFE2}), 
and $R_{\rm tot}/P_{\rm tot}\simeq0.053$ and $3000\,{\rm s}\times0.053\sim150$~s, 
where 3000~s is the average value of equation (\ref{eq:type1}).

\bsp	
\label{lastpage}
\end{document}